\documentclass[fleqn,usenatbib]{mnras}

\usepackage[T1]{fontenc}

\DeclareRobustCommand{\VAN}[3]{#2}
\let\VANthebibliography\thebibliography
\def\thebibliography{\DeclareRobustCommand{\VAN}[3]{##3}\VANthebibliography}

\usepackage{graphicx}	% Including Fig. files
\usepackage{amsmath}	% Advanced maths commands
\usepackage{tabularx}
\usepackage{amssymb}	% Extra maths symbols
\usepackage{orcidlink}
\usepackage{newtxtext,newtxmath}

\newcommand{\bs}{\boldsymbol}

\title[Radial migration in gas-rich discs]{Enhanced rates of stellar radial migration in gas-rich discs at high redshift}

\author[Zhang et al.]{HanYuan Zhang,$^{1}$\thanks{hz420@cam.ac.uk (HZ)}\orcidlink{0009-0005-6898-0927}
Thor Tepper-García,$^{2}$\orcidlink{0000-0002-1081-883X}
Vasily Belokurov,$^{1}$\orcidlink{0000-0002-0038-9584}
N. Wyn Evans,$^{1}$\orcidlink{0000-0002-5981-7360}
Takafumi Tsukui,$^3$\orcidlink{0000-0002-1499-6377}
\newauthor
Hillary Davis,$^{2}$\orcidlink{0009-0006-2670-0453}
Joss Bland-Hawthorn,$^{2}$\orcidlink{0000-0001-7516-4016}
Jason L. Sanders,$^{4}$\orcidlink{0000-0003-4593-6788}
Oscar Agertz$^{5}$\orcidlink{0000-0002-4287-1088}
\\
$^{1}$Institute of Astronomy, University of Cambridge, Madingley Road, Cambridge CB3 0HA, UK\\
$^{2}$Sydney Institute for Astronomy, School of Physics, University of Sydney, NSW 2006, Australia\\
$^{3}$Kavli Institute for the Physics and Mathematics of the Universe, 5 Chome-1-5 Kashiwanoha, Kashiwa, Chiba 277-8583, Japan\\
$^{4}$Department of Physics and Astronomy, University College London, London WC1E 6BT, UK\\
$^{5}$Lund Observatory, Division of Astrophysics, Department of Physics, Lund University, Box 118, SE-22100 Lund, Sweden
}

\date{Accepted XXX. Received YYY; in original form ZZZ}

\pubyear{2024}

\begin{document}
\label{firstpage}
\pagerange{\pageref{firstpage}--\pageref{lastpage}}
\maketitle

\begin{abstract}
Radial migration and dynamical heating redistribute stars within galactic discs and thereby modify the chemo-kinematic structure of their host galaxies. Usually, these secular processes are studied in N-body and hydrodynamical simulations of Milky Way analogues with stellar-dominated discs. In contrast, discs at high redshift are gas-rich, which may qualitatively change how secular evolution proceeds. We use the {\sc Nexus} framework to construct and evolve a suite of isolated galaxies with fixed halo and disc mass but varying initial disc gas fraction, from 0\% to 100\%. We show that in gas-rich models, the root-mean-square change in stellar angular momentum is up to a factor of two larger than in gas-poor analogues and is accompanied by stronger radial and vertical heating, leading to enhanced radial mixing. We further dissect the role of gas in specific migration channels. For bar-driven migration, corotation-resonance dragging dominates in gas-poor discs, whereas in gas-rich discs stars more readily reach and accumulate at the outer Lindblad resonance, which acts as a barrier. The high radial mixing efficiency in gas-rich phases can flatten the stellar metallicity gradient relative to that of the initial gaseous disc within only a few orbital timescales. Together, these results imply that radial mixing in early, gas-rich discs is substantially more vigorous than in late-time, gas-poor discs, naturally producing distinct evolutionary tracks for chemically bimodal discs such as that of the Milky Way.

\end{abstract}


%%%%%%%%%%%%%%%%%%%%%%%%%%%%%%%%%%%%%%%%%%%%%%%%%%

%%%%%%%%%%%%%%%%% BODY OF PAPER %%%%%%%%%%%%%%%%%%

\section{Introduction}

% Secular evolution dominates the late time evolution of the stellar disc

\begin{figure*}
    \centering
    \includegraphics[width=0.88\textwidth]{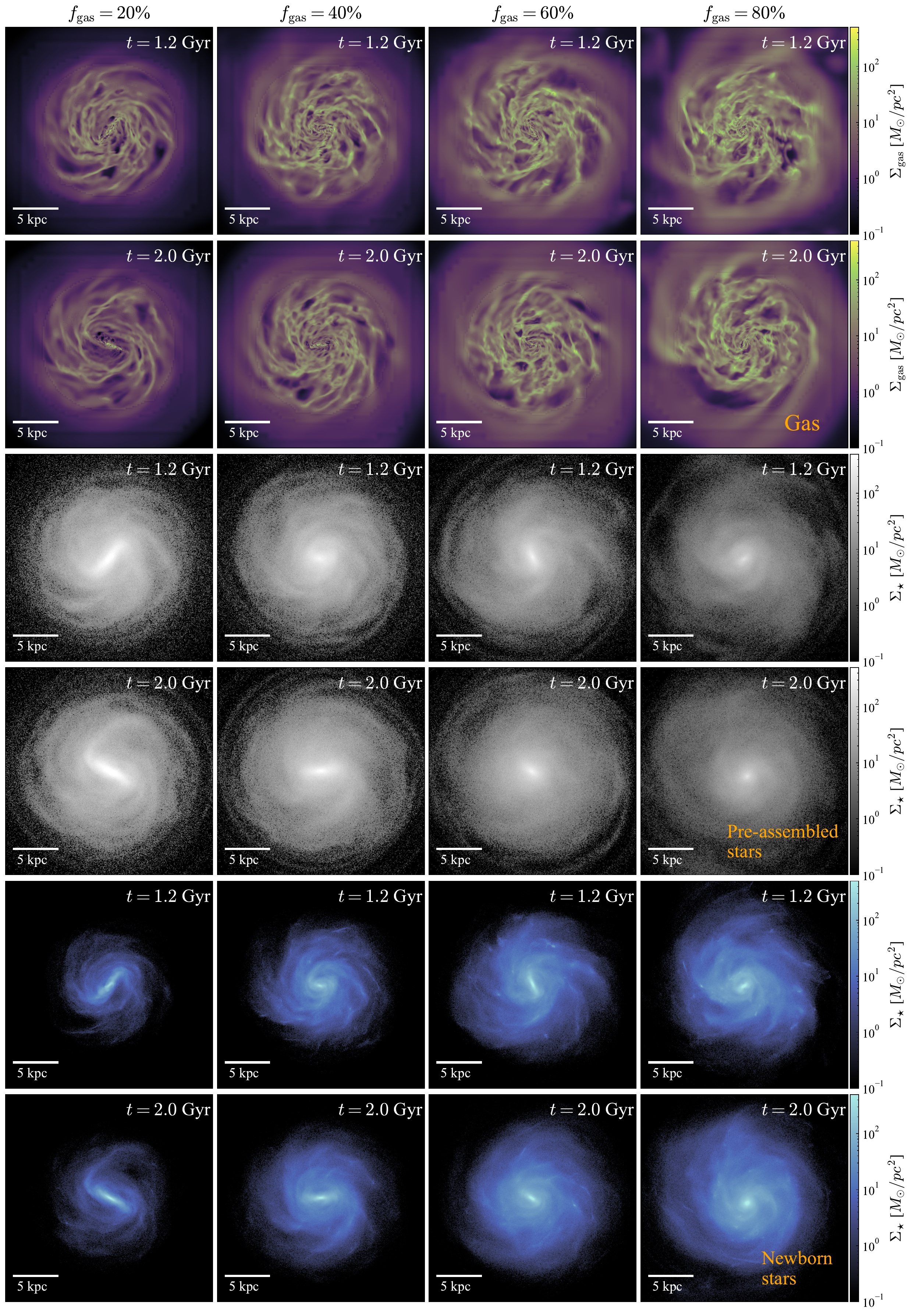}
    \caption{{\it Top sets:} the surface density maps of the gaseous component in the snapshots of galaxies at 1.2 Gyr and 2.0 Gyr (final snapshot) with turbulent gas in Set~1 of $f_{\rm gas}=20\%, 40\%, 60\%, 80\%$ from left to right. {\it Middle sets:} same as the top, but they show the surface density maps of the pre-assembled disc stars in the snapshots. {\it Bottom sets:} same as the top, but they show the surface density maps of the newborn disc stars in the snapshots. All the panels are 24~kpc across.}
    \label{fig:snap}
\end{figure*}

The Atacama Large Millimeter/sub-millimeter Array (ALMA) and the James Webb Space Telescope (JWST) have revealed kinematically cold stellar and gaseous discs at very early cosmic times, even at $z>3$ \citep[e.g.][]{Rizzo2020,Lelli2021,Tsukui_Iguchi2021,Ferreira2022,Ferreira2023, Rowland2024}. The dating of the Milky Way's thick disc formation time also indicates epochs around $z\approx2-3$ \citep[e.g.][]{Miglio2021,Queiroz2023,Belokurov2024}, which raises interesting questions regarding the early evolution of these ancient discs. 

Secular evolution, encompassing processes such as radial migration and dynamical heating, redistributes stars within disc galaxies and dominates their evolution at late cosmic times ($z\lesssim1$) \citep[e.g.][]{Sellwood_Binney2002, Roskar2008, Schonrich2010, Sellwood2014}. Radial migration is typically described as a change in the angular momentum of stars, whereas dynamical heating refers to changes in the radial and vertical actions of their orbits. Numerical simulations have extensively explored how stellar orbits varies, highlighting torques from spiral structure \citep[e.g.][]{Sellwood_Binney2002, Minchev2011_resonanceoverlap, Hamilton2024}, galactic bars \citep[e.g.][]{Halle2018, Khoperskov2020, Haywood2024, Zhang2025, Marques2025}, scattering by giant molecular clouds (GMCs), and interactions with external galaxies \citep[e.g.][]{Quillen2009, Bird2012, Carr2022, Hebrail2025}. Among these mechanisms, internal non-axisymmetric perturbations such as bars and spiral arms are generally regarded as one of the dominant sources of migration in discs. The same time-dependent perturbations also increase the random motions of stars, so that angular-momentum redistribution is typically accompanied by disc heating \citep[][]{Minchev2006, Daniel2019, Hamilton2024}. This secular re-arrangement of stellar orbits has important chemo-dynamical consequences: it broadens stellar metallicity distributions at fixed radius \citep{Minchev2013, Minchev2025, Kubryk2013}, modifies radial and vertical metallicity gradients \citep{Schonrich_MM2017, Kawata2017}, generates azimuthal abundance trends \citep{DiMatteo2013}, and may contribute to the disc's $\alpha$-bimodality \citep{Sharma2021_alphabimodality, Chen2023, Zhang2025}. Most of these studies, however, utilise test-particle or N-body simulations in which the disc is predominantly stellar. Consequently, their conclusions primarily describe the late-time evolution of discs ($z\lesssim1$), whereas early discs are known to be gas-dominated \citep{Tacconi2020}.

High-redshift observations have shown that disc substructures (e.g., bars and spiral arms) can emerge in the early gas-rich discs \citep[e.g.][]{Guo2023, Ferreira2023, Kuhn2024, Tsukui2024, EspejoSalcedo2025, LeConte2025}. As demonstrated in simulations, these non-axisymmetric perturbations are known to drive secular evolution in early discs, and can even lead to morphological 
% TTG: I feel 'late-type' and 'early-type' are outdated concepts; I'd rather say 'disc=dominated' and 'spheroidal', or 'rotationally=supported' vs. 'disperson-dominated', ...
transformations from rotation-supported to dispersion-dominated galaxies \citep{vanderWel_Meidt2025_EASE}. Despite our knowledge about secular evolution in gas-poor discs at the late time, the understanding of the impact of gas on radial migration in these gas-rich discs remains insufficient. Hydrodynamical/cosmological simulations have been employed to study radial migration \citep[e.g.][]{El-Badry2016, Okalidis2022, Haywood2024, Marques2025, Bernaldez2025}, and the impact from stellar-feedback is carefully analysed by \citet{El-Badry2016}. More recently, \citet{vanDonkelaar2025} investigated radial migration in early discs within GigaEris cosmological zoom-in simulations and reported interesting behaviour where stars born at very high redshift ($z\gtrsim6$) preferentially migrate inwards, forming a pseudo-bulge. This differs from the behaviour observed at stellar-dominated late times, which typically involves a mix of inward and outward migration \citep{Sellwood_Binney2002, Roskar2008}. 

The prevalence of turbulent gas, often due to elevated star-formation rates, could alter the formation and dissolution rates of galactic bars \citep{BH2024}, an analogy that could potentially be extended to other density perturbations such as spiral arms. Therefore, gas-rich galaxies at high redshift may provide a distinct environment for secular evolution compared to gas-poor discs at late times. Recent studies concerning the radial migration history of the Milky Way indeed reveal a distinct radial migration strength in its old thick disc and its young thin disc (\citealt{Minchev2013}, Zhang et al. in prep.). 
%To phenomenologically understand the impact of gas on the secular evolution of galaxies, we analyse the time evolution of stellar orbits using controlled isolated galaxy simulations within the {\sc Nexus} framework \citep[][]{TG24}.
% TTG edited:
% To isolate the role of gas in secular evolution, we use controlled isolated galaxy simulations built with the {\sc Nexus} framework \citep[][]{TG24}. In these numerical models, we follow the orbits of disc stars to quantify how secular processes depend on gas fraction.
To isolate the role of gas in secular evolution, we use a set of controlled, isolated galaxy simulations built with the {\sc Nexus} framework \citep[][]{TG24}. These simulations are ideal for the task, as the only initial difference between them is the initial gas fraction of the synthetic galaxy.

The paper is structured as follows. We introduce the adopted galaxy models in Section~\ref{section:sim}. We present the behaviour of radial migration and dynamical heating in galaxies versus the gas fraction in Section~\ref{sec:results}. In Section~\ref{sec:implication}, we link our results to the related studies of the Milky Way and external disc galaxies to discuss the implications, and we summarise the article in Section~\ref{sec:conclusion}.

\section{Simulations} \label{section:sim}

We analyse two sets of synthetic galaxy models presented by \citet{BH2024} and \citet{Davis2025}, respectively, both of which have been created within the {\sc Nexus} framework \citep{TG24}. {\sc Nexus} exploits the power of the Action-based GAlaxy Modelling Architecture \citep[{\sc agama};][]{Vasiliev2019} to create fully self-consistent initial conditions using a distribution-function (DF) based approach, which are evolved with a modified version of the Adaptive Mesh Refinement (AMR) N-body/hydrodynamical code {\sc Ramses} \citep{Teyssier2002}, which includes the method developed by \citet[][and references there in]{age21l} to account for galaxy formation physics.

In the following, we describe the features of these isolated galaxy models relevant to this work. Further details can be found in the aforementioned references.

\subsection{Set 1} \label{subsec::model1}

The first set of models we analysed in this work is a subset of the simulations originally presented by \citealt{BH2024}. It consists of a suite of synthetic galaxies with properties comparable to the MW's progenitor at redshift $z \approx 3$, originally referred to as `MW-progenitor analogues'. This suite helps to understand the evolution of Milky Way-analogue galaxies in their young, gas-rich phase.

Each of the synthetic galaxies in this set is modelled as a multi-component system consisting of a dark matter (DM) host halo with a mass $\sim10^{11} M_\odot$, a {\em pre-assembled} stellar disc, and a gaseous disc with a mass corresponding to a fraction between 0 and 100\% of the total disc mass (see below, Eq.~\ref{eq:fgas}). We refer to the stars in the pre-assembled disc as `pre-existing', in contrast to the stars that form out of the gas, dubbed `newborn'. The pre-assembled disc has initial Toomre $Q$ parameter of $\approx 2$, and scale height $h_z\approx0.15~\rm kpc$.

The disc mass fraction in our adopted set is $f_{\rm disc} = 0.5$, where
\begin{equation}
    f_{\rm disc} = \left(\frac{V_{c,\,{\rm disc}}(R)}{V_{c,\,{\rm tot}}(R)}\right)_{R = 2.2R_{\rm disc}},
\end{equation}
$R_{\rm disc}$ is the disc scalelength, and $V_c^2 = R~{\rm d}\Phi/{\rm d}R$ is the circular speed curve of the galaxy.

It is worth noting that values of $f_{\rm disc} \gtrsim 0.3$ render a galaxy unstable to bar formation within a Hubble time, with the timescale for bar formation, $\tau_{\rm bar}$, exponentially declining with $f_{\rm disc}$ \citep[the `Fujii relation';][]{fuj18a,BH2023}. This relation predicts that a galaxy with $f_{\rm disc} = 0.5$ shall form a bar within $\tau_{\rm bar} \sim2$~Gyr (but the galactic bar could be short-lived when $f_{\rm gas}$ is high, \citealt{BH2024}).

The only {\em initial} difference between the synthetic galaxies in our set is their initial gas fraction,
\begin{equation} \label{eq:fgas}
    f_{\rm gas} =  \left(\frac{M_{\rm disc,\,gas}}{M_{\rm disc}}\right),
\end{equation}
where $M_{\rm disc,\,gas}$ and $M_{\rm disc}$ are the gas disc mass and the total (gas + pre-assembled stars) disc mass, respectively. It takes on values $f_{\rm gas} \in \{20\%,\,40\%,\,60\%,\,80\%\}$, i.e. our simulation suite of consists of four synthetic galaxies, which are initially identical in every aspect except their $f_{\rm gas}$.
%JBH->HZ: we have 1% tracer stars in the 100% gas fraction case, so these can be studied.

All synthetic galaxies were evolved for $\sim2$~Gyr. Since the gas in these galaxies is `active', i.e. it is subject to heating and cooling, and is allowed to form stars, accounting for their feedback and enrichment, the gas becomes turbulent, and it is depleted over time, and does so at different rates in each galaxy (see Fig.~3 in \citealt{BH2024}). Therefore, the evolution of each galaxy is different, leading to different final states.\footnote{We kindly refer the reader to the latter study and \citet{TG24} for further details about these simulations.}

Fig.~\ref{fig:snap} displays the surface density distribution of the gas, of the pre-assembled stellar disc and of the newborn stars at two snapshots of each galaxy. It is worth noting that galaxies with $f_{\rm gas}=(20\%,\, 40\%)$ do feature a stellar bar at their centre at the final snapshot as expected from their $f_{\rm disc}$, while the more gas-rich models $f_{\rm gas}=(60\%,\, 80\%)$ do not. However, as explained in \citet{BH2024}, these galaxies with higher gas fraction do indeed form a bar, albeit short-lived, and it evolves into a bulge structure by the final snapshot.

As shown in the bottom rows of Fig.~\ref{fig:snap}, the surface density distributions of newborn stars depend on the gas fractions of the galaxies, as expected. A difference in the initial radius of stellar particles would affect their migration efficiency, as we will demonstrate in Section~\ref{sec:results}. Therefore, to ensure a fair comparison of the dependence of secular evolution on the gas fraction of the galaxy, we focus our analysis on the evolution of the pre-assembled stellar disc -- as opposed to the disc made out of newborn stars. For example, the effective radii of these stars, $R_e$, at the final snapshots all have values around $R_e\approx3.5-4$~kpc, very similar to each other. However, as we illustrated in Appendix~\ref{appendix::new_vs_old_stars}, the differences in radial migration behaviour of the pre-assembled disc stars and newborn stars at the same birth radius are minor.

%TTG->HZ: moved paragraphs below to Results
% We computed the orbital actions ($J_R$, $J_z$, $J_{\phi}$ or $L_z$) of these stars for simulation snapshots of every $\sim50$~Myr using \textsc{agama} \citep{Vasiliev2019}.

% We analyse the changes in orbital actions over $1.5$~Gyr since the first $500$~Myr years, on which we wait for the disc to settle. The effective radius, $R_e$, of these galaxies at the final snapshot are similar around $R_e\approx3.5-4$~kpc.

To assess the effect of the active, turbulent gas on the dynamics of the pre-assembled stellar disc, we additionally analysed a synthetic galaxy identical to the one with $f_{\rm gas}=20\%$, but evolved using a strict isothermal equation of state, i.e. without cooling/heating and no star formation; we refer to this as an `inert' gas system.

\subsection{Set 2} \label{subsec::model2}

Since the galaxies in Set 1 were evolved for a relatively short time, there is less time for the bar-driven migration to build up.

Hence, we consider a second set of synthetic galaxies to better understand the effect of gas on bar-driven radial migration over longer time scales. This set consists of two galaxies, a gas-free ($f_{\rm gas}=0$) galaxy, and a gas-bearing galaxy with $f_{\rm gas}=10$\%.

The base galaxy model consists of a DM host halo of mass $1.2\times10^{12}M_\odot$, a pre-assembled stellar bulge of mass $1.3\times10^{10}M_\odot$, and a pre-assembled stellar disc with mass $4.3\times10^{10}M_\odot$. The gas-bearing galaxy includes a gaseous disc with an initial mass of $4.6\times10^{9}M_\odot$. These system parameters are broadly consistent with the corresponding features of the MW \citep{BH_Gerhard2016}, except that the current gas fraction is closer to 10\% in the real MW. For further details about the galaxy setup and the simulation method, we refer the reader to \citet{Davis2025}.

Both galaxies were evolved for $\sim4$~Gyr. 
In both, a central stellar bar forms, and the bar strength, measured by the maximum amplitude of the quadrupole moment $A_2/A_0$, is comparable in both galaxies, at a level $A_2/A_0\sim0.45$.
%TTG-HZ: Perhaps include a plot showin A2/A0(t)? I don't think any of our earlier papers has shown A2 for these galaxies side-by-side.

%TTG-HZ: Moved to Results
% We calculate the quadrupole strength of the snapshots with cadence $50$~Myr, and label the bar formation time in these galaxies as the time that $A_2/A_0$ reaches $0.25$, which the same is used in \citet{Fragkoudi2025} (but see also other definitions, e.g. \citealt{BH2023}). The bar-formation times in these galaxies are $\sim1$~Gyr for the galaxy with $f_{\rm gas}=20\%$ and $\sim1.5$~Gyr for the gas-free galaxy, respectively. This allows us to inspect the bar-driven radial migration over $\gtrsim2.5$~Gyr since bar formation in these galaxies. We compute the pattern speed and corotation resonance (CR) radius and outer Lindbald resonance (OLR) radius in these galaxies since bar formation using the continuity equation with code developed in \citet{Dehnen2023} (see also Fig.~11 in \citealt{Davis2025}). This method calculates the pattern speed with uncertainty using a single snapshot information to high accuracy as verified in simulations, and has also been used to compute the bar pattern speed of the Milky Way and LMC \citep{Zhang2024_patternspeed, Araya2025} and shows good performance.

\section{Analysis and Results}\label{sec:results}

We analyse the time evolution of orbital properties of the pre-assembled stellar disc as a function of the gas fraction using the isolated galaxy models introduced above. 

\subsection{Analysis}

% Set 1
We study the migration signature in the galaxies from Set 1 based on the statistical analysis of the variation of stellar orbits using their orbital actions. To this end, we computed the three orbital actions\footnote{Their corresponding angle variables are of no relevance to us for the time being, and are therefore not considered.} ($J_R$, $J_z$, $J_{\phi}$ or $L_z$) of the pre-existing stars with a time cadence of $\sim50$~Myr using \textsc{agama} \citep{Vasiliev2019} with the axisymmetric gravitational potential of the corresponding snapshots. The gravitational potential are computed using the \texttt{CylSpline} method in \textsc{agama} for the stellar and gaseous disc, and \texttt{Multipole} method for the DM halo.

Given the initial `burst' of stellar activity right from the outset -- which is virtually unavoidable in these types of idealised models \citep[c.f.][]{BH2024} -- we allow the disc to settle for the first 500 Myr, and focus our analysis of the changes in all three orbital actions over the last $1.5$~Gyr of evolution.

Specifically, we calculate the root-mean-square (RMS) changes of the orbital actions of a stellar population at each epoch, which we denote as $\delta J_i$
\begin{equation}
     \delta J_i = \sqrt{\langle (J_i - J_{i,0})^2\rangle} ,
\end{equation}
where $i=(R, z, \phi)$, and $J_{i,0}$ is defined as the orbital actions at $t=500~\rm Myr$ throughout the paper unless otherwise specified. 

% Set 2
The analysis of the migration signature in the galaxies in Set 2 is based on the quadrupole strength of the bar, which is performed with a cadence of $50$~Myr over the time span of the simulation.

Following earlier work \citep{fuj18a},
% wrong ref: \citep[e.g.][see also \citealt{fuj18a}]{Fragkoudi2025},
%TTG-HZ: I think an earlier reference, e.g. by Athanassoula, is more appropriate; see if you can find one.
we mark the bar formation epoch as the time when the amplitude of the $m=2$ mode climbs to $A_2/A_0 \gtrsim 0.2$ \citep[for a more physically motivated definition, see][]{BH2023}. Based on this, we find that the bar-formation times are $\sim1$~Gyr for the galaxy with gas and $\sim1.5$~Gyr for the gas-free galaxy, respectively. This allows us to inspect the bar-driven radial migration over $\gtrsim2.5$~Gyr since bar formation in these galaxies. We compute the pattern speed and corotation resonance (CR) radius and outer Lindblad resonance (OLR) radius in these galaxies since bar formation using the continuity equation method \citep[][see also Fig.~11 in \citealt{Davis2025}]{Dehnen2023}. %This method calculates the pattern speed using a single snapshot as verified in simulations, and has also been used to compute the bar pattern speed of the Milky Way and LMC \citep{Zhang2024_patternspeed, Araya2025}.
%JBH: the problem here is when to start the clock - exponential fitting is better because T_o is then rigorously defined.
%HZ: I see the point, but I don't think it's necessary to have a very rigorously defined T_o for our purpose as we are not looking into the bar dynamics itself. So I took the simple approach, taking a consistent timescale for both galaxy that defines when does the bar become effective to the disc. 

% JBH: walter & thor & hillary find the code has problems... best not to say anything about performance.
% and shows good performance.

\subsection{Radial migration v.s. gas}\label{subsec::RMS_Lz}

\begin{figure}
    \centering
    \includegraphics[width=0.99\linewidth]{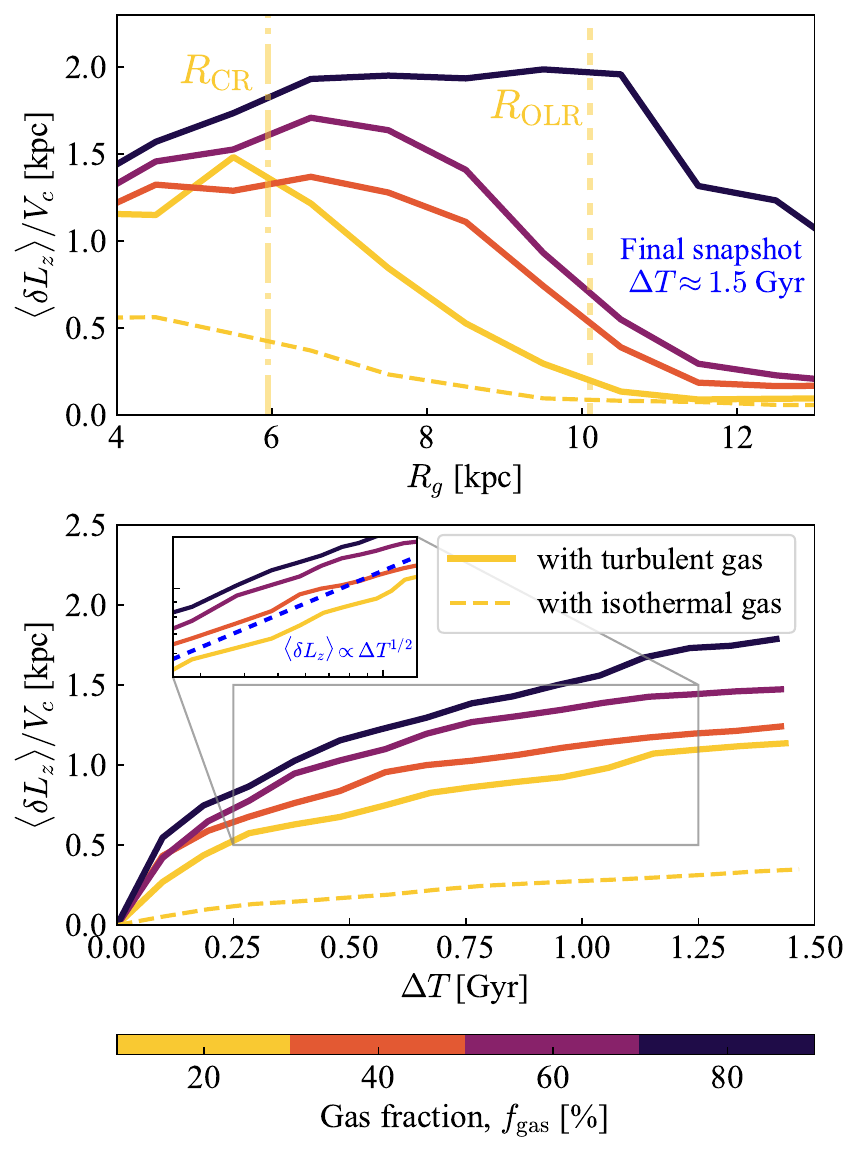}
    \caption{The root-mean-square variation of stellar orbital angular momentum normalised by the circular velocity of the galaxies. {\it Top:} the mean RMS variation of angular momentum as a function of the guiding radii of the stars in the final snapshot of the galaxies after $\Delta T \approx 1.5$~Gyr since the initial  ($t =500~\rm Myr$) snapshot. Each solid line is coloured according to the gas fraction of the galaxy, and the yellow dashed line is for the analogue galaxies with isothermal gas of $f_{\rm gas}=20\%$. The vertical dash-dotted line labels the current corotation radius of the galaxy with $f_{\rm gas}=20\%$. {\it Bottom:} the time evolution of the angular momentum variation traced in different snapshots since time $\Delta T$ after the $500~\rm Myr$ snapshot, averaging among stars with guiding radius between $4-13$~kpc. The zoomed-in panel shows the region in the log--log scale, and the blue dotted line is the $\langle\delta L_z\rangle\propto\Delta T^{1/2}$ line to indicate the power-law nature of the angular momentum variation in these galaxy models. A different choice of the guiding radius range does not qualitatively affect the trends.}
    \label{fig:RMS_Lz}
\end{figure}

We first examine the radial migration behaviour as a function of gas fraction by inspecting the RMS variation of angular momentum for galaxies in Set 1. In the upper panel of Fig.~\ref{fig:RMS_Lz}, we present the radial migration strength for stars at various guiding radii in the final snapshot ($\Delta T\approx1.5$~Gyr), expressed as $\delta L_z/V_c \approx \delta R_g$. This analysis is not performed for stars within the bar regions; thus, we exclude the inner $4$~kpc in all Set 1 galaxies for consistency. The angular momentum variation demonstrates a clear difference among galaxies with varying gas fractions, indicating that migration is much more efficient in gas-rich discs. In the lower panels of Fig.~\ref{fig:RMS_Lz}, we illustrate the time variation of the radial migration strength. We marginalise over stars within the final guiding radius range of $4-13$~kpc to obtain the overall radial migration strength at each snapshot. Angular momentum variation is consistently more vigorous in gas-rich discs at all times. On average, the variation in the angular momentum in the most gas-rich disc in this experiment ($f_{\rm gas}=80\%$) is approximately $1.5-2$ times greater than that of its most gas-poor analogue ($f_{\rm gas}=20\%$).

The $\delta L_z/V_c$ values for the galaxy with $f_{\rm gas}=20\%$ are presented with a yellow curve in Fig.~\ref{fig:RMS_Lz}. The solid curve corresponds to the case of star-forming, turbulent gas, while the dashed curve corresponds to the case of isothermal gas with constant temperature of $10^4~\rm K$. We find that radial migration is much weaker in galaxies with isothermal gas than in their turbulent-gas counterparts. This difference likely stems from the different density structure of the gas in either case (see further discussion in \citealt{TG25, Davis2025}).
% TTG: I 'dared' to edit the previous and following pars eventough I did not write them originally

A greater difference in the angular momentum variation is observed in the outer part of the disc, whereas the discrepancy is smaller at small radii, comparing the black line in the upper panel of Fig.~\ref{fig:RMS_Lz} to other lines. We attribute this to the stellar bar that forms in the gas-poor simulations: in particular, the $f_{\rm gas}=20\%$ galaxy hosts a strong, long-lived bar that has been acting on the inner disc for about $1$~Gyr by the time of the final snapshot. The formation and deceleration of a stellar bar can drive, or substantially amplify, radial migration in the inner disc through several mechanisms \citep[e.g.][]{Minchev_Famaey2010, Khoperskov2020, Haywood2024, Zhang2025, Marques2025}. As shown in the upper panel of Fig.~\ref{fig:RMS_Lz}, the angular momentum variation in the $f_{\rm gas}=20\%$ galaxy around the corotation resonance is comparable to its gas-rich analogues at that radius, but it becomes less efficient outside $R_{\rm CR}$ compared to the others. Our results suggest that angular momentum variation is stronger in gas-rich galaxies at radii outside the corotation resonances, yet the prevalence of a bar in gas-poor galaxies may lead to more complicated behaviour inside the corotation resonance.

Regarding the radial dependence of angular momentum variation, a rapid decrease is observed with increasing guiding radius in discs with $f_{\rm gas}\leq60\%$ as shown in the yellow, orange and purple lines in the upper panels of Fig.~\ref{fig:RMS_Lz}. Note that a similar behaviour has been noted in the Galactic disc using the age-metallicity distributions of stars (\citealt{Haywood2024}; Zhang et al. in prep.). The angular momentum variation in the galaxy with $f_{\rm gas}=80\%$ maintains a plateau up to $10$~kpc and then also begins to decline in the outer disc. %This phenomenon could be explained by a reduction in perturbations in the outer disc.

\subsubsection{Causes of the excess $L_z$ variation in the gas-rich discs}

\begin{figure}
    \centering
    \includegraphics[width=0.98\linewidth]{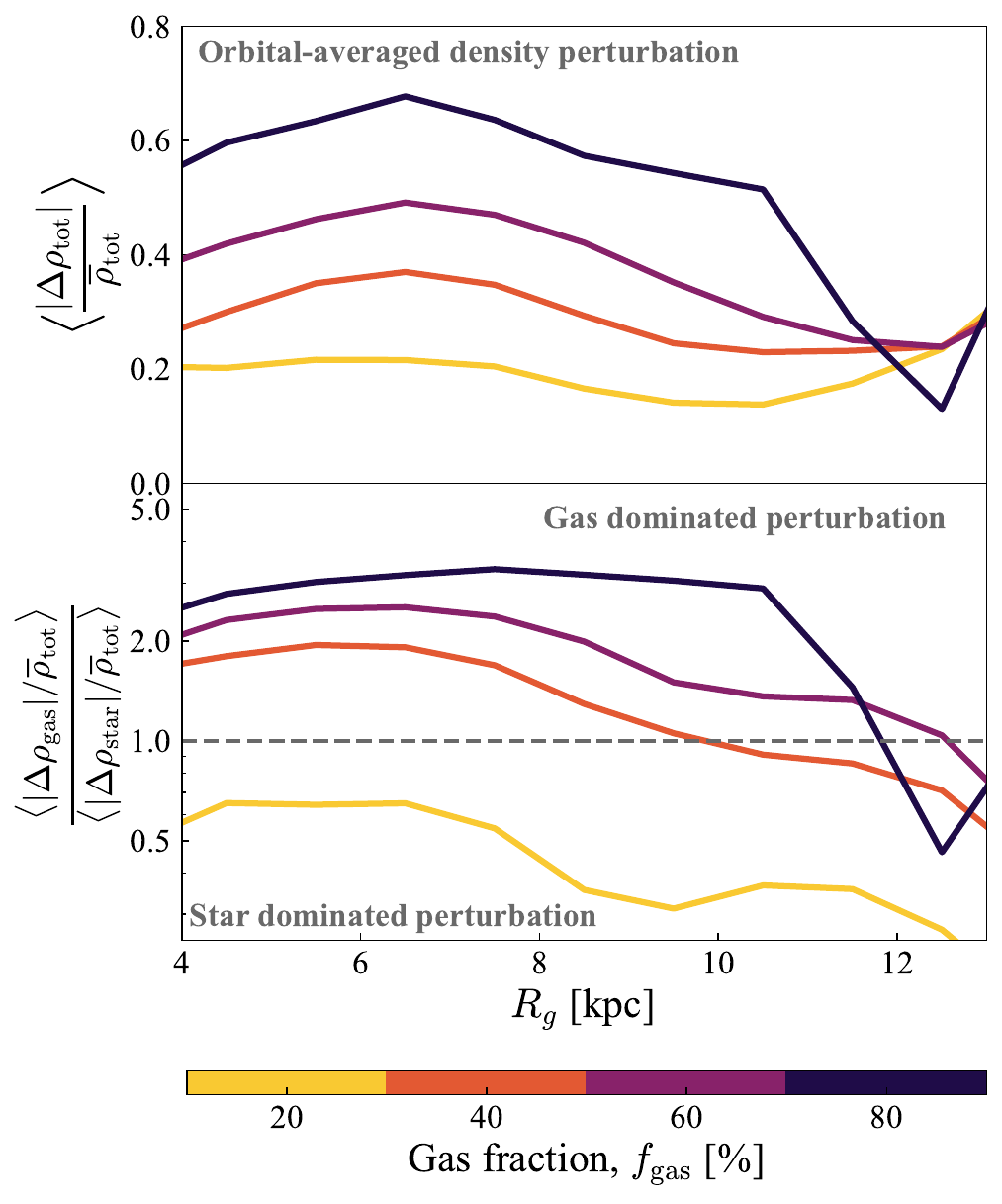}
    \caption{The mean relative density perturbation experienced by the stars over the $\sim1.5$~Gyr orbital times in the galaxies with different gas fractions, $f_{gas}$. {\it Top:} the mean total density perturbations averaged along the stellar orbits for stars in different guiding radii in the final snapshot. Each line is coloured according to the gas fraction of the galaxy. {\it Bottom:} the ratio of the relative density perturbation contributed by the gas and stars. The grey dashed line is the $\rm ratio=1$ line, for which regions above the line means that the gas component dominates the density perturbations, and regions below the line means that stars dominate the perturbations.}
    \label{fig:density_perturbation}
\end{figure}

As the non-axisymmetric perturbations in galactic discs are the main drivers of angular momentum variation and diffusion \citep{Romeo1990, Sellwood_Binney2002, BT08}, we compute the integrated density perturbation experienced by each stellar particle over the $1.5$~Gyr time span as follows:
\begin{align}
    \left\langle \frac{|\Delta \rho_i|}{\overline{\rho}_{\rm tot}} \right\rangle &= \frac{1}{T}\int_0^T \frac{|\Delta \rho_i ({\bs x(t)}, t)|}{\overline{\rho}_{\rm tot}(\bs x(t),t)} dt \approx \frac{1}{N_{\rm snap}}\sum_j^{N_{\rm snap}} \frac{|\Delta \rho_{i,j} ({\bs x_j})|}{\overline{\rho}_{{\rm tot},j}(\bs x_j)},
\end{align}
where $\Delta \rho_{i,j} ({\bs x_j})$ denotes the non-axisymmetric density perturbation of substance $i=\{\rm gas,\,star,\,tot\}$ at snapshot $j$, $\overline{\rho}_{{\rm tot},j}$ is the total baryon density averaged azimuthally at the star's annulus, and ${\bs x_j}$ represents the location of the star at snapshot $j$. The time-integrated density perturbation could also be considered as the orbital-averaged density perturbation and therefore could be correlated to the orbital-averaged diffusion coefficient of the actions \citep[see more in section 7.4 of][]{BT08}. In practice, $\Delta \rho_{i,j} ({\bs x_j})$ was calculated using a spatial cell of size $300^3$~pc$^3$. We calculated the total density perturbation and the perturbation from stars and gas individually. The radial trends of the total density perturbation, averaged along the orbits, are shown in the upper panel of Fig.~\ref{fig:density_perturbation}, which are further averaged among the stars occupying the same guiding radii bins. As the Figure demonstrates, stars in gas-rich discs experience stronger density perturbations throughout all $1.5$~Gyrs of their orbital history. Consequently, there are stronger non-axisymmetric perturbations available for stars in gas-rich discs to exchange angular momentum, leading to a greater diffusion of stellar orbital angular momentum.

We compare the relative contribution of gas and stellar content (including both pre-assembled and newborn stars) to the total density perturbation by examining the ratio of $\langle|\Delta \rho_{\rm gas}|/\overline{\rho}_{\rm tot}\rangle$ and $\langle|\Delta \rho_{\rm star}|/\overline{\rho}_{\rm tot}\rangle$ in the lower panel of Fig.~\ref{fig:density_perturbation}. It is only in the galaxy with a $20\%$ gas fraction that stars dominate the density fluctuations. For galaxies with $f_{\rm gas}\geq 40\%$, gas is the main contributor to the density perturbations, consistent with the more top-heavy cloud mass spectrum as $f_{\rm gas}$ increases \citep{fen21,fen23}.
This implies that in these galaxies, the gas behaviour primarily governs the secular evolution of the stellar disc \citep{BH2025}.

We decompose the density field of the four galaxies into azimuthal Fourier modes and find that more gas-rich discs have relatively stronger power in higher-$m$ modes, corresponding to finer, smaller-scale non-axisymmetric structure. This behaviour is expected, as gas tends to clump more effectively than stars. The strength of the long-wavelength quadrupole moment also fluctuates more intensely in gas-rich environments. These factors could further boost the angular momentum variation in gas-rich galaxies, though the detailed correlation between the wavelength of perturbations and the strength of radial migration remains unestablished and is attributed to future work.

Baryon sloshing, as discussed in \citet{BH2025}, could also enhance the variation of stellar angular momentum. \citet{BH2025} demonstrated that in gas-rich galaxies, the potential minimum of the baryons can deviate from that of the dark matter halo by up to a few kiloparsecs through a random walk, and the amplitude of this random walk increases with higher gas content. Therefore, in the rest frame of stars and gas, the dark matter halo, an inherently axisymmetric component, effectively provides additional non-axisymmetric and time-dependent perturbations to the stars. Consequently, the stronger time-varying potential of the dark matter halo, relative to the stellar component, can also induce changes in the orbital actions of stars.

%HZ: This discussions for the causes of stronger radial migration in the gas-rich discs are added

\subsection{Heating-to-migration ratio, ${\cal H}$}\label{subsec:JRz_Lz}
%JBH: seems important to me, shall we give it a symbol? if so, use H everywhere in text, no need to change labelling on figs.
%HZ: I like that symbole

\begin{figure*}
    \centering
    \includegraphics[width=0.99\linewidth]{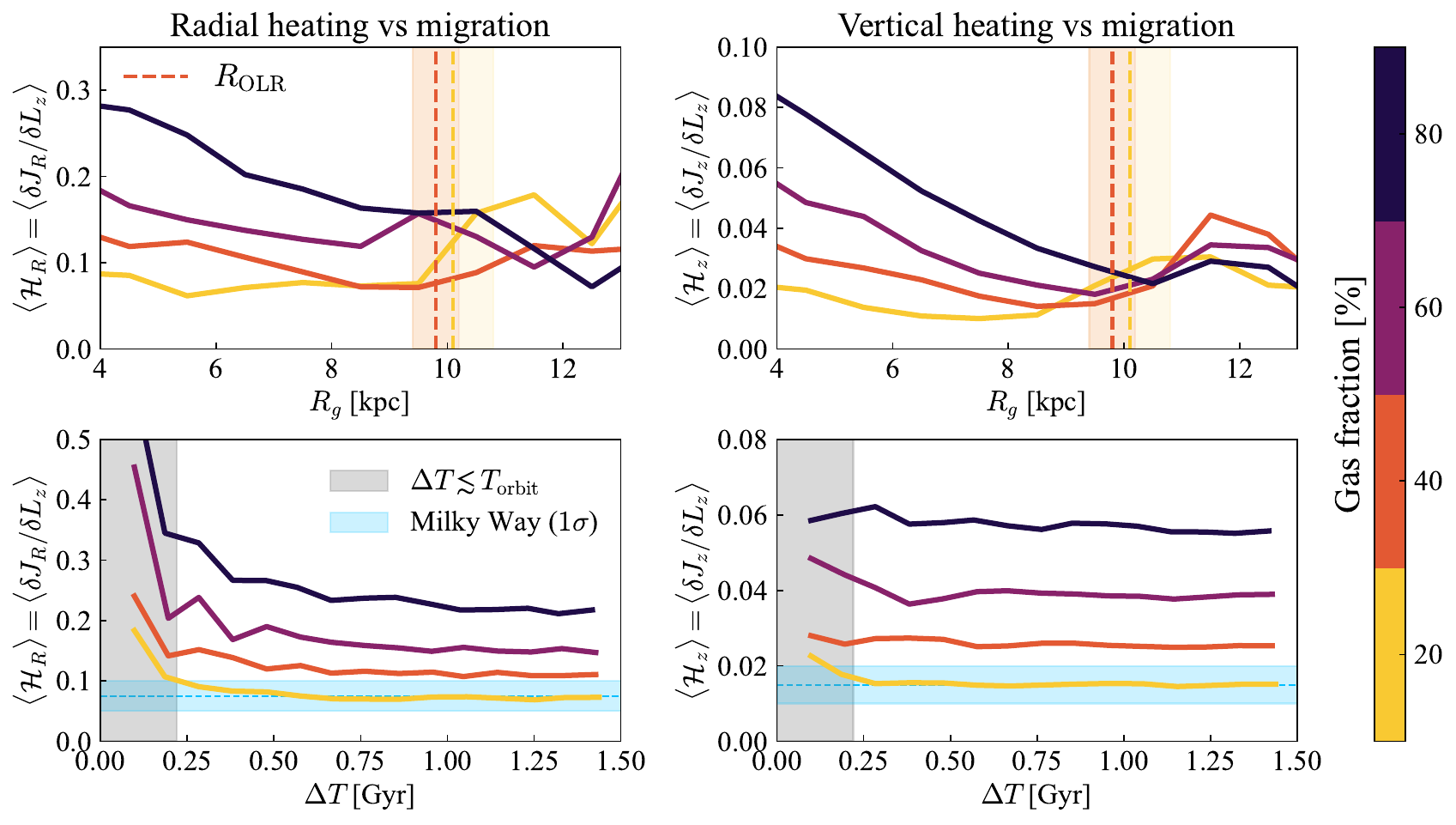}
    \caption{Similar to Fig.~\ref{fig:RMS_Lz} but shows the radial and time dependence of the averaged radial and vertical heating-to-migration ratio of galaxies with different gas fractions. The heating-to-migration ratios, $\mathcal{H}$, are quantified as the ratio of the RMS variation of radial and vertical actions to the RMS variation of the azimuthal actions, $\mathcal{H}_{R,z} =\delta J_{R,z}/\delta L_z$, which the radial heating-to-migration ratio is shown on the left, and the vertical one is on the right. {\it Top:} the mean heating-to-migration ratios of stars as a function of the guiding radii in the final snapshot after $\sim1.5$~Gyr since the beginning. The vertical dashed lines and the bands are the current OLR radius of the galaxy with $f_{\rm gas}=20\%, 40\%$ and its $1\sigma$ region, which roughly corresponds to the radius for which the radial trends of $\delta J_{R,z}/\delta L_z$ change. {\it Bottom:} the time dependence of the mean heating-to-migration ratios, which we averaged over the stars between $4<R_g/\rm kpc<13$ in each snapshot. The grey shaded regions on the left denote the time that $\Delta T$ is less than an orbital period. The horizontal blue  shaded regions denote the measured heating-to-migration ratio of the Milky Way (\citealt{Frankel2020}; Zhang et al. in prep.)}
    \label{fig:heating_migration}
\end{figure*}

We evaluate the heating accompanying angular momentum diffusion by calculating the ratio of the RMS changes in radial and vertical actions to that in the angular momentum, ${{\cal H}_{R, z}}=\delta J_{R,z}/\delta L_z$. This quantity is also adopted by \citet{Frankel2020, Hamilton2024} and Zhang et al. (in prep.) for the same purpose. Intuitively, ${\cal H}$ quantifies the radial and vertical variation of the orbits compared to the expansion of the orbital size. Similarly to the analysis presented above, we illustrate the radial trends of the radial and vertical heating-to-migration ratios in the upper two panels of Fig.~\ref{fig:heating_migration} and their time dependence in the lower panels, obtained by averaging over the guiding radii between $4-13$~kpc for each snapshot with a cadence of $\sim50$~Myr.

As the Figure demonstrates, both radial and vertical heating-to-migration ratios vary systematically with gas fraction: gas-rich discs show higher ratios, i.e. stronger heating for a given amount of migration. Clearly, the ratio ${\cal H}$ can vary quite significantly as a function of the gas fraction. For instance, in the galaxy with $f_{\rm gas}=20\%$, there is relatively little heating compared to migration, with ${\cal H}\approx 0.07$, whereas for $f_{\rm gas}=80\%$, this increases to ${\cal H} \approx 0.24$. 

The dimensionless quantity ${\cal H}$ is useful for understanding the detailed secular evolution processes within a galaxy. For dynamical interactions associated with a non-axisymmetric perturbation of a constant pattern speed $\Omega_p$, the ratio of the change in radial and azimuthal actions in the epicyclic limit is given by \citep[][]{Sellwood_Binney2002}:
\begin{equation}
    \Delta J_R=\frac{\Omega_p-\Omega}{\kappa}\Delta L_z,
    \label{eqn:heating_migration}
\end{equation}
where $\Omega$ and $\kappa$ are the azimuthal and radial frequencies of the orbit. Consequently, migration associated with the corotation resonances of the perturbation ($\Omega_p \approx \Omega$) can lead to $\Delta J_R \approx 0$ for arbitrary $\Delta L_z$, while secular evolution involving other resonances causes finite heating during angular momentum drifting. Non-linear effects, particularly when resonances of perturbations with different pattern speeds overlap, can also stochastically affect stellar orbits, leading to an additional change in radial actions as angular momentum drifts \citep[e.g.][]{Minchev2011_resonanceoverlap, Daniel2019}. Therefore, the increase in ${\cal H}_R$ with increase $f_{\rm gas}$ suggests that interactions with the corotation resonances of bars and spirals are more important for gas-poor galaxies, while the other resonances and interactions becomes more significant in gas-rich galaxies. We further examine this in Appendix~\ref{appendix::spectrogram} by calculating the spectrograms of the galaxies in Set 1. Indeed, we find that corotation structures are more prominent in the $f_{\rm gas}=20\%$ galaxies. Lindblad resonances begin to appear with increasing $f_{\rm gas}$, and eventually the spectrogram becomes irregular for the $f_{\rm gas}=80\%$ galaxy (see Fig.~\ref{appendix:fig:spectrogram}). We will study the correlation between a galaxy's spectrograms and its secular evolution outcomes more closely in future works.

In the upper panels of Fig.~\ref{fig:heating_migration}, the heating-to-migration ratio decreases with increasing radius, though this trend reverses in the outer discs. In galaxies with $f_{\rm gas}=20\%$ and $40\%$, the radius at which these trends reverse coincides with the Outer Lindblad Resonance (OLR) radii, $R_{\rm OLR}$, of the galactic bar. This suggests that the reversal could be associated with the OLR radial migration barrier \citep[][]{Halle2015}, where bar-induced radial migration diminishes around the bar's OLR radius, and angular momentum exchange becomes minimal outside the OLR. This leads to more stochastic orbital secular evolution, thereby increasing the heating-to-migration ratio. Alternatively, it could simply be due to insufficient material in the outer disc to form non-axisymmetric perturbations capable of supporting regular secular evolution, resulting in more stochastic dynamical interactions.
%JBH: i find the above discussion much more plausible that Minchev's claim a decade ago that migration GROWS outer discs which i never understood.
% HZ: great

Interestingly, the radial and vertical heating-to-migration ratios of these isolated galaxies do not depend on time over timescales longer than an orbital period, i.e., when $\Delta T \gg T_{\rm orbit}$. Therefore, the behaviour we reported above is robust with respect to the simulation time and the snapshot adopted. Recent measurements of the heating-to-migration ratio for the Milky Way's low-$\alpha$ disc show strong consistency with the simulation results for the galaxy with $f_{\rm gas}=20\%$, yielding ${{\cal H}_{R,\,\rm MW}}\approx0.075 \pm 0.025$ and ${{\cal H}_{z,\,\rm MW}}\approx0.015 \pm 0.05$ (\citealt{Frankel2020}, Zhang et al, in prep). Zhang et al. (in prep.) also observe that the ratios become higher outside $R_{\rm OLR, \,MW}\approx11$~kpc \citep[e.g.][]{Zhang2024_patternspeed, Dillamore2025a}, reaching approximately ${\cal H}_{\rm R,\, MW}\sim0.13$ and ${\cal H}_{\rm z,\, MW}\sim0.04$, respectively, which is consistent with our simulated findings. All these suggest that MW's low-$\alpha$ disc is consistent with a gas-poor disc, but caution should be taken when comparing the Milky Way's disc to discs in isolated galaxy simulations.

% \hanyuan{In echo to the recent debates about why the Milky Way's disc is so cold, and here we provide a perspective that gas fraction in a galaxy is important.}

% \hanyuan{The heating-to-migration ratios generally declines with increasing radius, but starts to increase outside the OLR}

% \hanyuan{Heating-to-migration ratio is approximately constant with time in these isolated disc, which the Milky Way's low-alpha young disc is consistent with the most gas-poor galaxy in this analysis}

\subsection{Effects on bar-driven migration}\label{subsec::bar-driven_RM}

\begin{figure}
    \centering
    \includegraphics[width=0.99\linewidth]{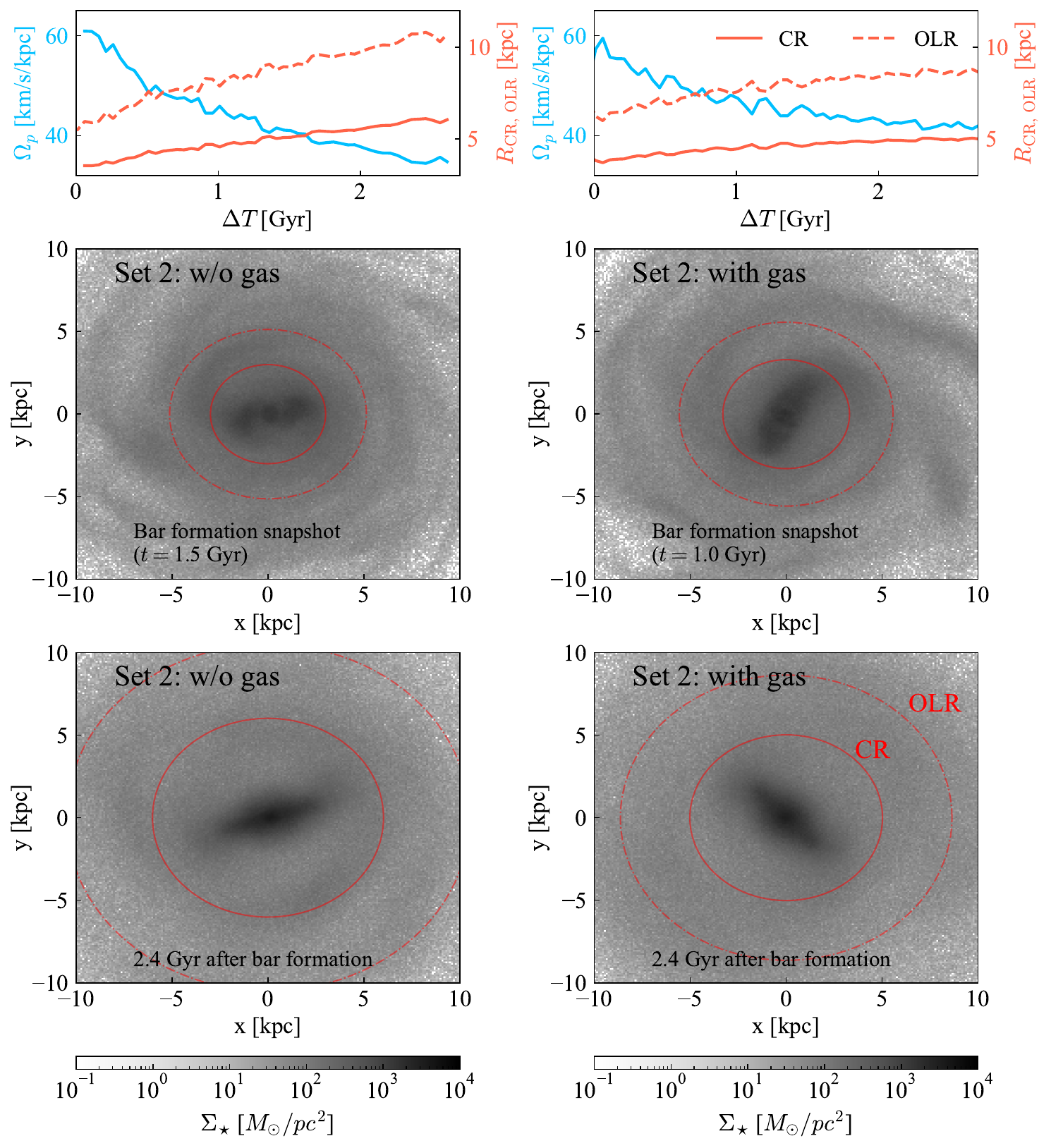}
    \caption{{\it Top:} the time evolution of pattern speed, $\Omega_p$, corotation and OLR radius of the two galaxies in Set 2. The pattern speed evolution is shown in blue. The corotation and OLR radius evolution are shown in red with solid and dashed lines, respectively. {\it Middle \& Bottom:} The stellar surface density maps of simulation snapshots of the two galaxies in Set~2 at the bar formation moment and 2.4~Gyr after that. The galaxy without gas is shown on the left and the galaxy with $f_{\rm gas}=10\%$ is shown on the right. The red solid and dot-dashed circle denote the CR and OLR radius of the corresponding snapshots. The expansion of the CR and OLR radius from the bar formation to the final snapshot is due to the bar deceleration.}
    \label{fig:snapshot_set2}
\end{figure}

\begin{figure*}
    \centering
    \includegraphics[width=0.99\linewidth]{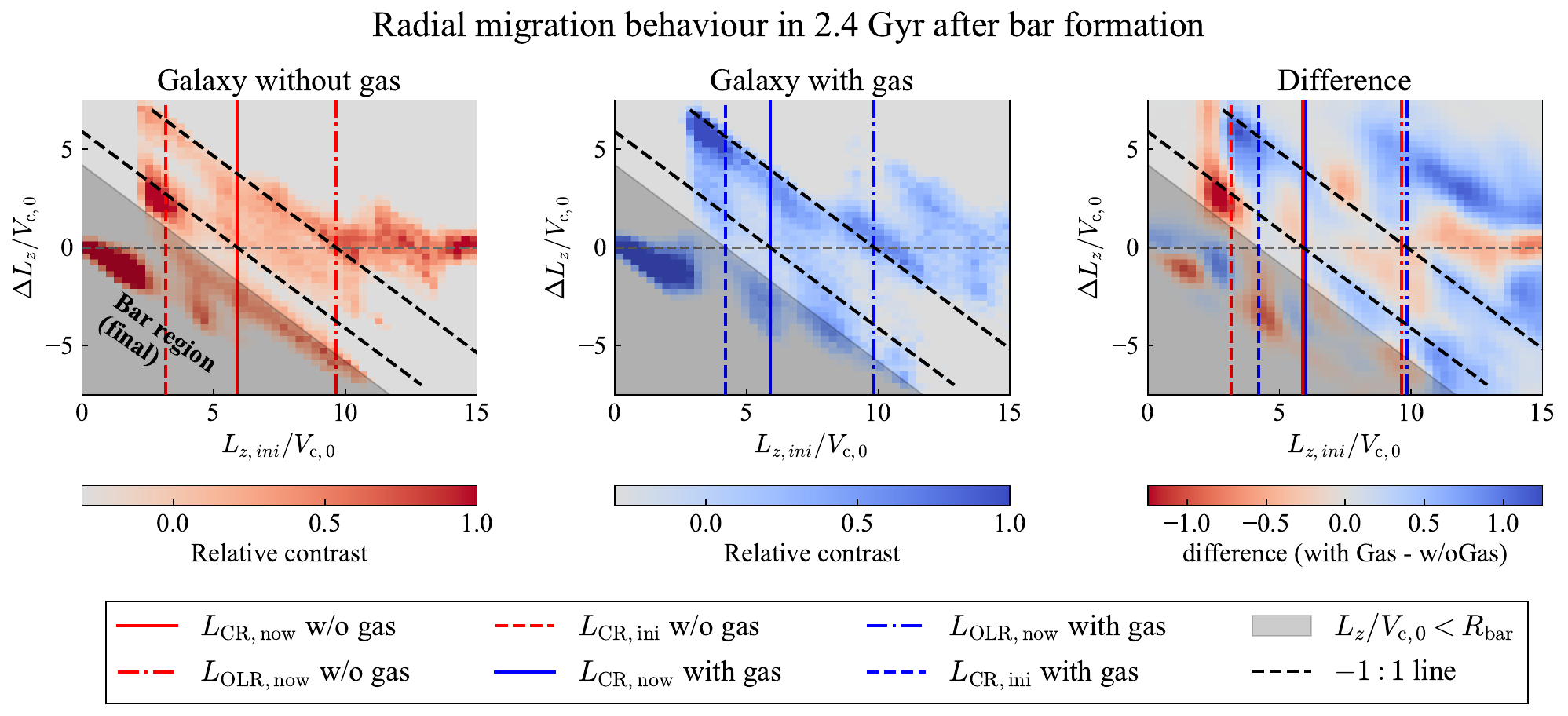}
    \caption{The unsharp initial angular momentum ($L_{z,\,\rm ini}$) v.s. the changes in angular momentum ($\Delta L_{z}$) plane of the two galaxies in Set~2 that have $f_{\rm gas}=0\%,\,10\%$ after $\sim2.4$~Gyr since the bar formation. The un-sharp masking is done by subtracting the Gaussian convolved $L_{z,\,\rm ini}-\Delta L_{z}$ plane with $\sigma$ of 5. {\it Left} shows this plane for the galaxy without gas. The grey shaded region in the lower left corner labels the stars that are currently in the bar region, for which $L_{z}/V_{c,0}<R_{\rm bar}$. The red dashed, solid, and dash-dotted lines correspond to the initial bar corotation angular momentum, final bar corotation angular momentum, and the final bar OLR angular momentum, respectively. The black-dashed lines show the $-1:1$ lines which join the grey $\Delta L_{z}=0$ line at the corotation and OLR line. Stars along these black dashed lines reside on the current corotation and OLR zone in the final snapshots. {\it Middle:} the same as the left panel but for the galaxy with gas. All the lines represent the same but in blue. {\it Right:} the contrast map that shows the difference between the left and middle panel. The red regions mean that the radial migration in the gas-free galaxy is more prominent in these regions, whereas the blue regions means that they are dominated by the galaxy with gas. The galaxies are scaled so that the present corotation angular momentum (solid lines) is the same.}
    \label{fig:delta_Lz_Lzini}
\end{figure*}

To examine how gas affects the detailed behaviour of radial migration, we use the second set of galaxy models, Set~2, introduced in Section~\ref{subsec::model2}. Compared to Set 1, galaxies in this set have a similar $f_{\rm disc}$ but are more massive overall, with $V_c \approx 215$~km/s. This model set comprises of two simulated galaxies: one that is gas-free and another with $10\%$ gas. Both galaxies host a central stellar bar, and both bars are decelerating, though the one in the gas-free galaxy decelerates more due to the absence of gas \citep[see e.g.][]{Beane2023, Semczuk2024}, as shown in the top panels of Fig.~\ref{fig:snapshot_set2}. The bar strength in both galaxies are similar to each other during the time interval we studied. Our primary aim is to understand the difference in bar-driven radial migration between galaxies with and without gas. Therefore, we treat the snapshot at the moment of bar formation as the initial snapshot and analyse the radial migration history in these galaxies since that point. The stellar surface density maps of the two galaxies at the bar formation and the final snapshot are shown in Fig.~\ref{fig:snapshot_set2}.

We use the $L_{z,\,\rm ini}~{\rm vs} ~\Delta L_z$ plane to analyse radial migration processes, where $L_{z,\,\rm ini}$ is the initial angular momentum at the moment of bar formation, and $\Delta L_z$ is the change in angular momentum \citep[e.g.][]{Sellwood_Binney2002, Minchev_Famaey2010, Halle2015}. Fig.~\ref{fig:delta_Lz_Lzini} shows the $L_{z,\,\rm ini}-\Delta L_z$ plane for the two galaxies in Set 2, approximately $2.4$~Gyr after bar formation. The left and middle panels of Fig.~\ref{fig:delta_Lz_Lzini} display the unsharp-masked $L_{z,\,\rm ini}-\Delta L_z$ plane for the gas-rich and gas-free galaxies, respectively, highlighting features of radial migration. The right panel illustrates the difference between these two; more precisely, we subtract the left panel from the middle one. Since the bar pattern speeds of these two galaxies differ at this snapshot, meaning their CR and OLR radii are also different, we rescale $L_{z,\rm ini}$ and $L_z$ of the gas-rich galaxy so that the corotation resonance of both galaxies matches, ensuring a fair comparison of bar-driven radial migration. The initial and final corotation angular momentum, and the final OLR angular momentum, are indicated by the dashed, solid, and dot-dashed vertical lines in all three panels. The black dashed lines, joining the final $L_{\rm CR}$ and $L_{\rm OLR}$ along the $\Delta L_z=0$ line, imply that stars occupying regions around these black dashed lines are located around the CR and OLR zones in the current snapshot after migration. The dark-grey shaded region represents stars currently inside the bar regions ($L_z/V_c<R_{\rm bar}$). Within this area, we observe several clumps and ridges corresponding to stars trapped by the bar during its growth. We focus on the area outside the dark-grey shaded region, as a discussion of the differences in bar structures between these two galaxies is beyond the scope of this paper.

In the left and middle panels, we identify two common ridges aligned with the two black dashed lines ($-1:1$ lines), which correspond to the radial migration driven by the bar's corotation and OLR. The red high-density clump in the left and right panels, located at the intersection of the red dashed line and the first black dashed line (i.e., $L_{z,\,\rm ini}\approx L_{\rm CR,\,ini}$ and $\Delta L_z\approx2.5V_{c,0}$), represents stars that have migrated due to bar corotation-resonance dragging \citep[e.g.][]{Khoperskov2020, ChibaSchonrich_2021, Zhang2025}. These stars were trapped by the bar's corotation resonances when the bar formed and subsequently migrated outwards with the expansion of the corotation radius during bar deceleration. This mechanism is more pronounced in the gas-free galaxy than in the gaseous galaxy. 

In contrast, the blue ridge along the second black dashed line, corresponding to the bar OLR resonance, is a more prominent feature in the middle and right panels. This ridge is caused by the same OLR migration barrier effect discussed previously, where stars initially inside the OLR at bar formation migrated outwards but were halted and accumulated at the OLR barrier \citep[][]{Halle2015}. \citet{Haywood2024}, using an isolated galaxy simulation, similarly illustrated that stars inside the initial corotation resonances of the bar at its formation migrated violently outwards during bar formation, with migration stopping at the OLR. This explains the blue over-density clumps in the middle panel at $L_{z,\,\rm ini}\lesssim L_{\rm CR,\,ini}$ and $\Delta L_z\approx5V_{c,0}$. This result suggests that corotation resonances become less significant for gas-rich galaxies, leading to stronger radial migration where stars more easily reach the OLR. This could be attributed to the additional density perturbations from the gas, which kick stars out of the corotation resonances during bar deceleration, allowing them to migrate further to reach the OLR. This difference between gas-free and gas-rich galaxies also explains why bar corotation resonance dragging is not frequently observed in galaxies from cosmological simulations, as those galaxies are prevalently gas-rich, and thus the role of the bar corotation is suppressed by turbulent gas. 

It is also noteworthy that the ridges we identified in the simulations are aligned with the black dashed lines with a slope of $-1$ instead of $-2$ as, for example, in Fig. 4 of \citet{Sellwood_Binney2002}. The difference is that the migration in these two galaxies in Set 2 is mainly driven by the galactic bar through the expansion of the corotation radius and the OLR barrier, whereas in \citet{Sellwood_Binney2002}, the ridges with slope $-2$ are a classical signature of stars in the horseshoe orbits trapped in corotation resonance that are swapping in and out around corotation.

Finally, as shown in the right panel of Fig.~\ref{fig:delta_Lz_Lzini}, the radial migration is also stronger for stars outside the OLR in the galaxy with gas. This behaviour is likely not associated with the galactic bar, and it is consistent with the results we discussed in Section~\ref{subsec::RMS_Lz}, for which the radial migration strength increases with increasing gas fraction.

% \hanyuan{1. corotation resonance dragging is stronger in the gas-free galaxy}

% \hanyuan{2. OLR is more prominent in the gas-rich disc}

% \hanyuan{3. Stronger migration signatures outside the OLR for more gas-rich galaxies}

\section{Implications: the Milky Way and early discs}\label{sec:implication}

\begin{figure}
    \centering
    \includegraphics[width=0.99\linewidth]{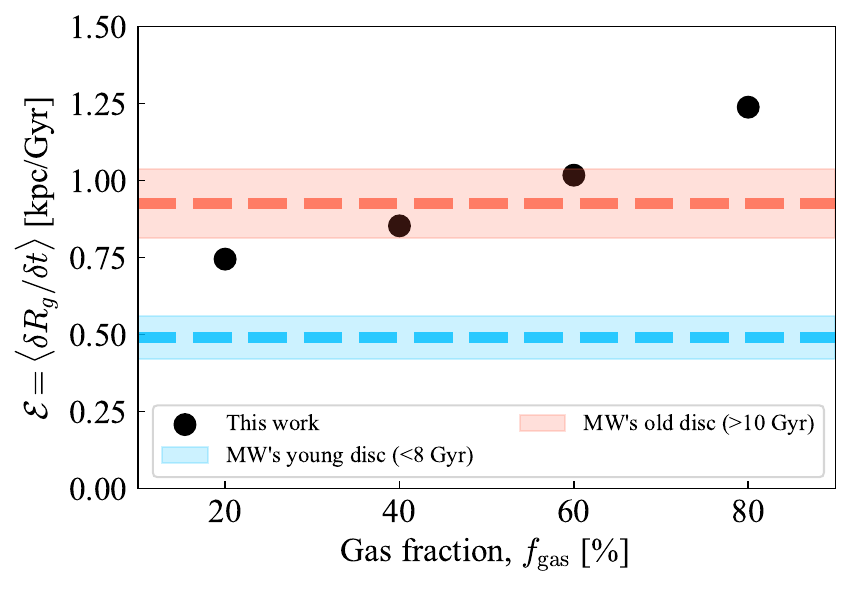}
    \caption{The mean ``radial migration efficiency'' ${\cal E}$ in terms of the rate of guiding radius variation in the four galaxies as a function of the gas fraction. The results in this paper of the four galaxies with turbulent gas in Set~1 are shown in the black dots, for which the gas-rich galaxies has stronger radial migration strength. The blue-shaded and the red-shaded regions are the measured time-averaged radial migration efficiency in the Milky Way's young, low-$\alpha$ and old, high-$\alpha$ discs with $1\sigma$ uncertainties. The two values of ${\cal E}$ of the Milky Way are computed by averaging $\delta R_g/\delta t$ over $2-8$~Gyr and $10-13$~Gyr, respectively. Note that this figure is to illustrate the correlation between the transition of radial migration efficiency in the Milky Way compared to the variation of gas fractions, but it should not be interpreted numerically as shown. }
    \label{fig:DRg_Dt}
\end{figure}

\subsection{Comparing to secular evolution in the Milky Way}

The distinct differences in radial migration strength and the heating-to-migration ratio observed across galaxies with varying gas fractions offer a unique avenue for probing the Milky Way's gas fraction evolution through its secular dynamics.

The secular evolutionary history of the Milky Way can be modelled using signatures within its age-metallicity-$R_g$ distributions, where the radial mixing of stars broadens the age-metallicity spread in a given radial annulus. \citet{Frankel2020} utilised red clump stars as tracers to quantify the radial migration efficiency and radial heating-to-migration ratio of the Milky Way's low-$\alpha$ disc. They derived $R_g\sim2.55~\rm kpc\sqrt{\tau/6~\rm Gyr}$ (assuming a constant angular momentum diffusion coefficient) and $\delta J_R/\delta L_z\approx 0.1$, values broadly consistent with the galaxies in Set 1 having $f_{\rm gas}=20\%~\rm or~40\%$. More recently, Zhang et al. (in prep.) performed a similar non-parametric modelling using subgiant/main-sequence turn-off stars in the solar neighbourhood to measure the same quantities. They obtained a time-averaged 
``radial migration efficiency''  ${\cal E}= \langle \delta R_g/\delta t\rangle\approx0.49\pm0.05 ~\rm kpc/Gyr$ for the low-$\alpha$ disc and  ${\cal E}\approx0.92\pm0.11 ~\rm kpc/Gyr$ for the high-$\alpha$ disc, as depicted by the blue-shaded and red-shaded regions in Fig.~\ref{fig:DRg_Dt} (we refer the reader to Zhang et al. in prep. for the limitation of this model). The corresponding  ${\cal E}$ for galaxies in Set 1, marked by black dots, suggests that the Milky Way's low-$\alpha$ disc (blue-shaded band) aligns more closely with the most gas-poor analogues in Set 1. This observed transition from high radial migration efficiency in the early high-$\alpha$ disc to lower efficiency in the later low-$\alpha$ disc could well be a consequence of a gas reservoir transition, implying that the Galaxy became gas-poor following the formation of the low-$\alpha$ disc. 

% JBH: if you are giving errors, never use \sim. you are claiming 3 sigma!!

The measured heating-to-migration ratios for the Milky Way's low-$\alpha$ disc, ${\cal H}_{\rm R,\,MW}\approx 0.075\pm 0.025$ and ${\cal H}_{\rm z,\,MW}\approx 0.015\pm 0.005$, are also consistent with the $f_{\rm gas}=20\%$ galaxy in Set 1, as further discussed in Section~\ref{subsec:JRz_Lz}. It is crucial to acknowledge that direct numerical comparisons between the Milky Way's disc and isolated galaxy simulations may be biased, particularly for the high-$\alpha$ disc, given the Milky Way's merger history \citep[e.g.][]{Belokurov2018, Helmi2018}. Thus, the values presented in Fig.~\ref{fig:DRg_Dt} should not be over-interpreted.

Furthermore, the Milky Way's age-metallicity distribution exhibits more structure than expected from a purely stochastic radial migration scenario \citep[][]{Haywood2024, Zhang2025, Cerqui2025}. Specifically, a "V-shaped" age-metallicity distribution is observed in the Milky Way's low-$\alpha$ disc around the solar neighbourhood \citep[][]{Xiang_Rix2022}. \citet{Zhang2025} interpreted these V-shaped bimodal sequences as resulting from the coexistence of stars born in the inner Galaxy that migrated outwards due to the expanding bar corotation radius during its deceleration, and stars born locally in the solar neighbourhood. This scenario is further supported by the radial dependence of the age-metallicity distribution and observed azimuthal chemical patterns \citep[][]{Khoperskov2025b}. Such findings underscore the significant role of bar corotation resonances in the Milky Way's radial migration history, corroborated by the low $\delta J_R/\delta L_z$ ratio measured. As discussed in Section~\ref{subsec::bar-driven_RM}, the influence of bar corotation resonances would be suppressed in a gas-rich disc due to additional perturbations from the gas. Consequently, for bar corotation resonance dragging to have dominated radial migration and reshaped the age-metallicity pattern in the low-$\alpha$ disc, our Galaxy must have remained gas-poor for over $6$~Gyr, for which the V-shaped sequences start to emerge.

\subsection{Metallicity gradients in the early discs}

\begin{figure}
    \centering
    \includegraphics[width=0.99\linewidth]{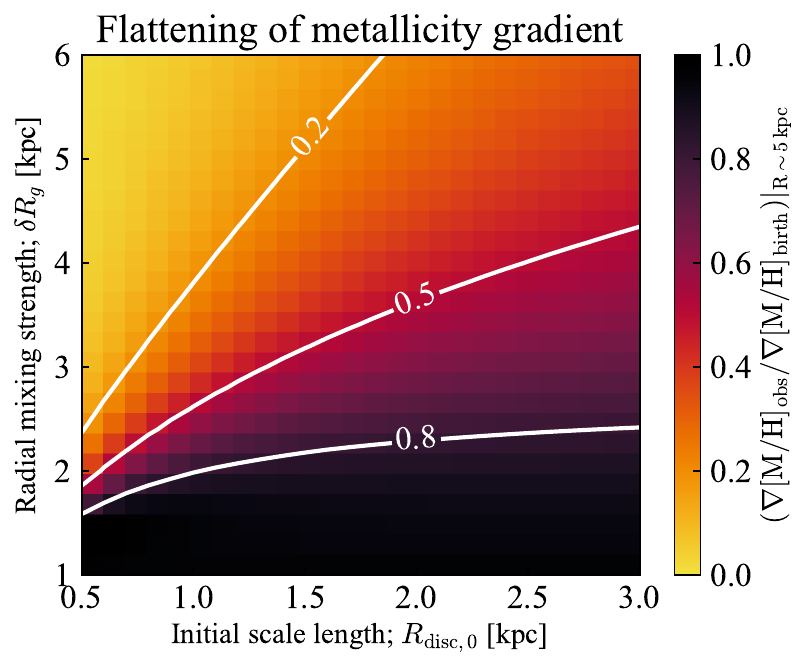}
    \caption{The expected flattening in radial metallicity gradient at a given radial mixing strength and initial disc scalelength, in terms of the ratio between the final and initial metallicity gradient. The lighter colour means a more significant flattening. The white lines are the contours when ratio is $0.2$, $0.5$, and $0.8$.}
    \label{fig:flatten_Z_grad}
\end{figure}

Secular evolution plays a pivotal role in the evolution of disc galaxies and their present-day observables \citep[e.g.][]{Sellwood_Binney2002, Schonrich_MM2017, Loebman2016, Bernaldez2025, Minchev2013, Minchev2025, Renaud2025}. Crucially, radial migration alters a galaxy's chemical pattern by broadening the metallicity distribution at a given age and radius, and simultaneously flattening the radial metallicity gradient. According to our results that the gas-rich galaxies experienced stronger radial mixing, we can expect a flatter metallicity gradient for the stellar population born in the gas-rich phase of the galaxy, which is typical for the high-$z$ discs \citep[][and references therein]{Tacconi2020}. Recent JWST and ALMA observations have also revealed detailed structures within high-$z$ stellar and gaseous discs, including numerous non-axisymmetric features such as multi-armed spirals and bars \citep[e.g.][]{Hodge2019, Meidt2023, Tsukui2024, LeConte2024, EspejoSalcedo2025}. Such structures provide a suitable environment for radial migration and heating, thereby accelerating secular evolution in early galaxies \citep[][]{vanderWel_Meidt2025_EASE}. 
We present a back-of-the-envelope calculation here for the expected flattening of the metallicity gradient given the disc profile and the amount of radial mixing. 

To describe the evolution of the metallicity gradient, the final mean metallicity at different radii can be calculated as:
\begin{equation}
    \langle {[\rm M/H]}\rangle|_{R_g} = \int [{\rm M/H}]p([{\rm M/H}]|R_g)~{\rm d} [{\rm M/H}],
    \label{eqn:mean_mh}
\end{equation}
where $p([{\rm M/H}]|R_g)$ is the final metallicity distribution at a given guiding radius. This final metallicity distribution, $p([{\rm M/H}]|R_g)$, can be expressed as:
\begin{equation}
    p([{\rm M/H}]|R_g) = \frac{\int p([{\rm M/H}]|R_b)p(R_g|R_b)p(R_b)~{\rm d}R_b}{\int p(R_g|R_b)p(R_b)~{\rm d}R_b},
    \label{eqn:final_mh_dist}
\end{equation}
where $R_b$ represents the stellar birth radius, which is marginalised over the entire disc as it is not directly observable. $p([{\rm M/H}]|R_b)$ is the initial metallicity distribution at the birth radius, $p(R_g|R_b)$ is the radial mixing model, and $p(R_b)$ is the birth radii distribution. We adopt the same assumptions as in \citet{Frankel2018, Frankel2020} (also employed by \citealt{Sharma2021_alphabimodality} and Zhang et al. in prep.). More precisely, our assumptions are as follows: 1) stars form in an exponential disc; 2) stars forming at the same time and radius have similar metallicities, and a linear metallicity gradient exists in the gas disc; and 3) angular momentum exchange occurs stochastically. These assumptions allow us to define: a) the initial metallicity distribution as
\begin{equation}
    p([{\rm M/H}]|R_b) \approx \delta([{\rm M/H}] - \overline{[{\rm M/H}]}|R_b),
    \label{eqn:initial_mh_dist}
\end{equation}
where $\overline{[{\rm M/H}]} = [{\rm M/H}]_0 + R_b\nabla[{\rm M/H}]_{\rm birth}$, with $[{\rm M/H}]_0$ being the central metallicity and $\nabla[{\rm M/H}]_{\rm birth}$ the metallicity gradient at birth, and $\delta$ means the Dirac delta function. b) For the radial migration model, we employ the solution to the 1D angular momentum diffusion equation, designed to approximately preserve an exponential disc profile well beyond the scalelength \citep{Sanders_Binney2015}:
\begin{equation}
    p(R_g|R_b) = A\times \mathcal{N}\left(R_g\,|\,R_b - \frac{\sigma_{R_g}^2}{2R_{{\rm disc},0}}, \,\sigma_{R_g}\right),
\end{equation}
where $A$ is a normalisation factor ensuring $\int_0^{\infty} p(R_g|R_b)\,dR_g=1$, $\sigma_{R_g}$ represents the strength of radial mixing, and $R_{{\rm disc},0}$ is the initial scalelength of the disc at the time of stellar formation. $\sigma_{R_g}$ is approximately $\delta R_g$ in the zeroth order, so to simplify the discussion, we will replace $\sigma_{R_g}$ by $\delta R_g$. c) The distribution of stellar birth radii is given by
\begin{equation}
    p(R_b) = \frac{R_b}{R_{\rm disc,0}^2}\exp\left(-\frac{R_b}{R_{\rm disc,0}}\right).
\end{equation}
Substituting Eq.~\ref{eqn:final_mh_dist} and \ref{eqn:initial_mh_dist} into Eq.~\ref{eqn:mean_mh}, we obtain:
\begin{equation}
\begin{split}
         \langle {[\rm M/H]}\rangle|_{R_g}& - [{\rm M/H}]_0 =\\
         &\frac{\int_0^{\infty}(R_b\nabla[{\rm M/H}]_{\rm birth})~p(R_g|R_b)~p(R_b)~dR_b}{\int_0^{\infty} p(R_g|R_b)p(R_b)~dR_b}.
\end{split}
\label{eqn:mean_mh_final}
\end{equation}
The crucial parameters in Eq.~\ref{eqn:mean_mh_final} are the amount of radial mixing $\delta{R_g}$ and the initial scale length $R_{{\rm disc},~0}$. We numerically evaluate Eq.~\ref{eqn:mean_mh_final} on a grid of $\delta{R_g}$ and $R_{{\rm disc},~0}$ around $R_g=5$~kpc to derive the gradient at this specific radius. As the final metallicity gradient is not necessarily constant with radius, we use $R_g=5$~kpc as a representative location. Fig.~\ref{fig:flatten_Z_grad} illustrates the ratio of the final and initial metallicity gradient, with lighter colours indicating a flatter final metallicity gradient. If we extrapolate the time evolution of the angular momentum variation as shown in Fig.~\ref{fig:RMS_Lz} to 6~Gyr ($z\approx1$), $\delta R_g$ could reach $3.5-4$~kpc for the galaxy with $f_{\rm gas}\geq60\%$, which could flatten the metallicity gradient to $0.2-0.5$ of its gradient at birth.  %The results demonstrate that for $\sigma_{R_g}\gtrsim4$~kpc, the metallicity gradient will flatten by half, regardless of the initial scale length, and that radial mixing exhibits a stronger effect for more compact discs.

%Observations of high-redshift galaxies predominantly suggest that early galaxies were gas-rich \citep[][and reference therein]{Tacconi2020}. When combined with the simulation results presented herein, this implies higher radial mixing rates in those early, gas-rich discs. 

High-redshift gaseous discs exhibit high turbulent speeds \citep[e.g.][]{ForsterSchreiber2009, Genzel2011, Wisnioski2015}. These turbulent kinematics are subsequently transferred to the initial kinematics of stars forming within these discs \citep[][]{vanDonkelaar2022, BH2025}, leading to discs that are "born hot" or heated in a very short timescale. These observations, alongside the discovery of young thick discs and the downsizing of discs \citep[][]{Elmegreen2017, Lian_Luo2024, Tsukui2025}, underscore the significant role of gas in driving the formation of bimodal discs. In this scenario, due to differences in gas content during the formation of the thin and thick discs, the efficiency of secular evolution might vary between them. Specifically, according to our results, radial mixing in the gas-rich, thick disc could be around twice as efficient as in the gas-poor, thin disc. Drawing on our calculation for the flattening of the metallicity gradient above, this suggests a flatter radial metallicity gradient in the thick disc compared to the thin disc, assuming that the gas-phase metallicity gradient does not vary significantly over cosmic time, as appears to be the case for the Milky Way \citep[][Zhang et al. in prep.]{Lu2024, Ratcliffe2025}, which could be an outcome of equilibrium behaviour \citep[][]{Johnson2025}. This offers a future pathway to verify the efficiency of early-time radial mixing through JWST observations and potentially with the GECKOS survey \citep[an edge-on disc galaxies survey, which can measure radial and vertical metallicity distributions for MW-sized galaxies;][]{vandeSande2024_GECKOS}. Intriguingly, \citet{Kawata2018} investigated the correlation between vertical and radial metallicity gradients in a simulated thick disc and its progenitor. They found that radial mixing necessitates a flat or positive radial metallicity gradient to account for the negative vertical gradient observed in the Milky Way's thick disc. Therefore, observations of vertical metallicity distributions in low-redshift thick discs could also provide crucial constraints on metallicity distributions in high-redshift gaseous discs, given the potentially strong radial mixing experienced by early thick discs.

%This is consistent with the observed evolution of the metallicity gradient over cosmic time \citep[][]{Li2025}, for which it was steep at high-z ($z\gtrsim4$) and flattens to be around zero at cosmic noon ($z\approx2$) and steeps again there. \citet{Li2025} interpreted this as a consequence of a different efficiency of gas mixing, but it could also be consistent with a different efficiency of stellar radial mixing. The steep metallicity gradient at the formation time was flattened by radial mixing of stars, and with decreasing radial mixing rate at later epoch due to the decrease of gas fraction allows the birth metallicity gradient to persist at lower redshift. Further observations of gas-phase metallicity at cosmic noon is needed to distinguish between these scenarios.

\section{Conclusions}\label{sec:conclusion}

We analysed the secular evolution of disc galaxies with various gas mass fractions and gas properties using controlled simulations within the {\sc Nexus} framework \citep[][]{TG24}. We computed the secular changes of stellar orbits through their orbital actions and examined the statistical variation of these actions over multiple orbital timescales. We found that stars in more gas-rich discs experienced more significant radial mixing. In general, the angular momentum variation in the galaxy with $f_{\rm gas}=80\%$ is almost twice as great as that in the galaxy with $f_{\rm gas}=20\%$. When examining the radial trends of angular momentum variation, we observed that the angular momentum of stars vary less significantly with an increasing radius at large radii ($R\gtrsim 2R_e$). However, these decreasing trends are weaker in gas-rich galaxies. Consequently, angular momentum variation in the outer disc is even more than twice as strong in gas-rich discs compared to gas-poor discs. We also calculated the heating-to-migration ratio, defined as the ratio between the variations of radial/vertical actions and angular momentum. Both radial and vertical heating-to-migration ratios are higher in gas-rich galaxies, indicating that orbits are heated faster in these galaxies, which further enhances radial mixing.

Non-axisymmetric perturbations are the main engine of secular evolution \citep[][]{Sellwood_Binney2002, Roskar2008}. We therefore computed the averaged non-axisymmetric density perturbation experienced by each stellar particle in the simulations across the investigated time span. We observed that stars in gas-rich galaxies experience more significant density perturbations, which can thus explain the aforementioned results. By decomposing the total density perturbation into contributions from stars and gas, we found that gaseous components are the main source of density perturbation for galaxies with $f_{\rm gas}\geq40\%$, whereas stars dominate the perturbation only in the galaxy with $f_{\rm gas}=20\%$. This suggests that a gaseous disc with cold kinematics provides more suitable environments for the formation of density perturbations, and that gas behaviour dominates secular evolution in gas-rich galaxies. The low value of the radial heating-to-migration ratio in gas-poor discs highlights the importance of corotation resonances of the perturbations in the secular evolution processes in those galaxies, whereas secular evolution in gas-rich discs is more stochastic.

We used bar-driven migration as an example to demonstrate the impact of gas on the detailed behaviours of radial migration. We inspected the angular momentum changes and the initial angular momentum at the moment of bar formation to reveal the detailed radial migration mechanisms in two barred galaxies, one with and one without gas. In both galaxies, the main features of radial migration are associated with the corotation resonances and the OLR of the galactic bars. The key difference is that the corotation resonance effect dominates in the gas-free galaxy, while the OLR has a stronger effect on the galaxy with gas. In the gas-free galaxy, stars around the bar's corotation radius initially migrate outwards, following the expansion of the corotation radius due to the bar deceleration, and subsequently reside around the final corotation radius \citep[see][]{Khoperskov2020, Zhang2025}. The same effect also occurred in the disc with gas, though it was much weaker. In comparison, stars in the gas-rich galaxy migrated further, reaching and accumulating around the final OLR due to the OLR migration barrier \citep{Halle2015}. This could be because stars are more easily removed from bar corotation resonances by the additional density perturbations contributed by the gas in gas-rich discs. These results again demonstrate the importance of corotation resonances of the substructures in the discs, particularly in gas-poor galaxies.

Finally, we discussed the implications of our results for the Milky Way's disc and for early high-redshift discs. We compared our numerical simulation results with existing secular evolution models of the Milky Way. Many studies suggest that the corotation resonances of bars and spirals are crucial for understanding the age-metallicity pattern of the low-$\alpha$ disc, and that this disc exhibits much less efficient radial migration compared to the high-$\alpha$ disc \citep[e.g.][Zhang et al. in prep.]{Frankel2018, Frankel2020, Hamilton2024, Cerqui2025}. According to our results, these observations in the Milky Way could hint at a transition from gas-rich to gas-poor disc phases when the low-$\alpha$ disc formed. To illustrate how significantly secular evolution can reshape the chemical pattern in discs, we presented an idealised calculation for the expected flattening of the metallicity gradient as a function of initial disc scale length and radial mixing strength. We argue that after efficient radial mixing in a gas-rich disc ($f_{\rm gas}\geq60\%$) for a few billion years, the stellar metallicity gradient can be flattened by more than half of its initial value (the gas-phase metallicity gradient). With a small initial disc scale length (i.e., a compact disc at formation), the metallicity gradient could even be flattened to less than $1/5$ of its original value. The bar and multi-armed spiral perturbations observed at high redshift could support this efficient secular evolution occurring early in a gas-rich environment \citep{Hodge2019, EspejoSalcedo2025, vanderWel_Meidt2025_EASE}. This could motivate future observations of the radial metallicity distributions of thick discs at both high and low redshifts to reveal radial mixing history at early time.
%This provides an alternative way to explain the evolution of the observed metallicity gradient over cosmic time \citep{Li2025}, and motivates further observations on the gas-phase metallicity gradient at different redshift. This could also possibly be examined with the metallicity gradient of the thick discs in the external galaxies locally to observe whether the thick disc with no metallicity gradient, like the Milky Way, is common. 

% The last numbered section should briefly summarise what has been done, and describe
% the final conclusions which the authors draw from their work.

\section*{Acknowledgements}

We acknowledge inspiring discussions with Mor Rozner, Neige Frankel, and Daisuke Kawata.

HZ thanks the Science and Technology Facilities Council (STFC) for a PhD studentship (grant number 2888170). 
TTG acknowledges partial financial support from the Australian Research Council (ARC) through Australian Laureate Fellowships awarded to JBH (FL140100278) from the School of Physics, University of Sydney, and partial funding through the James Arthur Pollock memorial fund awarded to the School of Physics, University of Sydney.
VB and NWE acknowledge support from the Leverhulme Research Project Grant RPG-2021-205: "The Faint Universe Made Visible with Machine Learning". 
TT is supported by the JSPS Grant-in-Aid for Research Activity Start-up (25K23392) and the JSPS Core-to-Core Program (JPJSCCA20210003).
JLS acknowledge support from the Royal Society (URF\textbackslash R1\textbackslash191555; URF\textbackslash R\textbackslash 241030). OA acknowledges support from the Knut and Alice Wallenberg Foundation, the Swedish National Space Agency (SNSA Dnr 2023-00164), and the LMK foundation.

% The Acknowledgements section is not numbered. Here you can thank helpful
% colleagues, acknowledge funding agencies, telescopes and facilities used etc.

%%%%%%%%%%%%%%%%%%%%%%%%%%%%%%%%%%%%%%%%%%%%%%%%%%



\section*{Data Availability}

The data and results of this paper could be shared upon reasonable request to the corresponding author of the paper. 

\bibliographystyle{mnras}
\bibliography{bibliography} % if your bibtex file is called example.bib

%\begin{thebibliography}{99}
%\end{thebibliography}

%%%%%%%%%%%%%%%%%%%%%%%%%%%%%%%%%%%%%%%%%%%%%%%%%%

%%%%%%%%%%%%%%%%% APPENDICES %%%%%%%%%%%%%%%%%%%%%

\appendix

\section{Fourier analysis and spectrograms}\label{appendix::spectrogram}

\begin{figure*}
    \includegraphics[width = 0.98\textwidth]{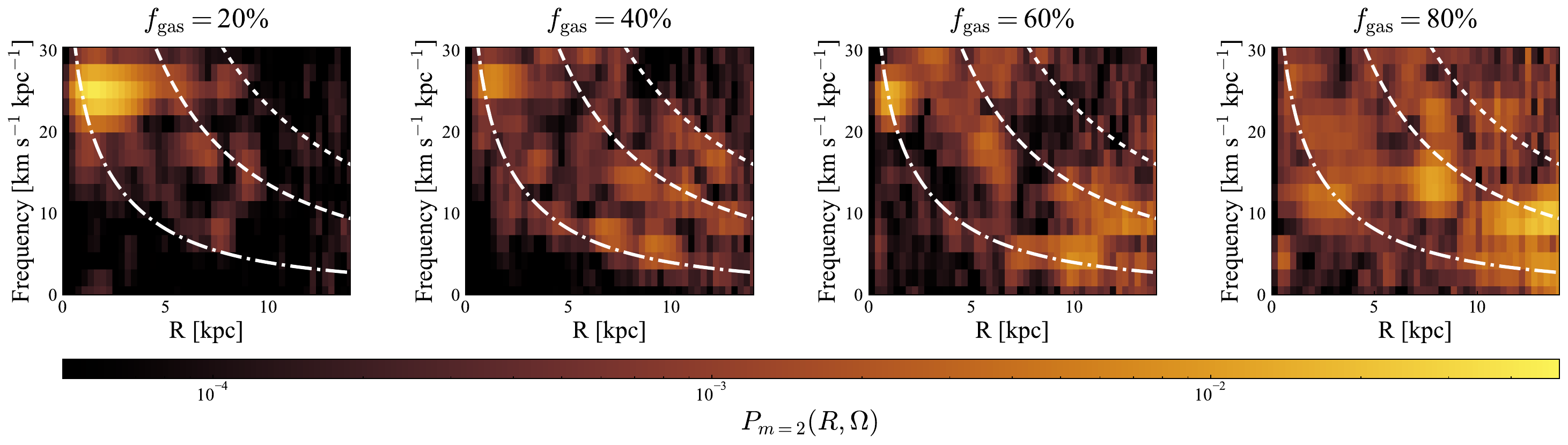}
    \caption{The spectrogram of the four galaxies in Set~1 with active gas with fraction $f_{\rm gas}=20\%,\,40\%,\,60\%,\,80\%$ from left to right. The white dash-dotted, dashed, and dotted lines in the panels represents the inner Lindbald resonance (ILR), corotation resonances, and outer Lindblad resonances (OLR) radius in these galaxies with the corresponding pattern speed. The corotation feature dominates the galaxy with $f_{\rm gas}=20\%$, but in more gas-rich galaxies, perturbations around ILR and OLR starts to emerge. }
    \label{appendix:fig:spectrogram}
\end{figure*}

To understand the reason that causes the difference in the orbital diffusion of the four galaxy models in Set~1, we compute the spectrograms of these galaxies \citep{Sellwood1986} using the same methodology and the window function as in \citet{Roskar2012}. We briefly summary the approach below. 

We calculate the Fourier moments of the galaxy at all snapshot that has a time cadence of $50~\rm Myr$ in the $1.5~\rm Gyr$ time interval we considered by 
\begin{equation}
    c_m(R) = \frac{1}{M(R)} \sum_{j=1}^{N} m_j\, e^{i m \phi_j}\,, 
\end{equation}
where $M(R)$ is the total baryon mass in the radial annulus with a width of $0.3~\rm kpc$. We focus on the quadrupole ($m=2$) moment as we are mainly interested in the impact from the bar/spiral structures, but the formula will remain general. We then compute the windowed discrete Fourier transform of the $c_m$ time series by
\begin{equation}
    C_{k,m}(r) = \sum_{j=0}^{S-1} c_{m,j}(r)\, w_j \, e^{-\,2\pi i\, j k / S}\,,
\qquad k=0,\ldots,S-1 ,
\end{equation}
where $S$ is the total number of snapshots, which $S=30$ in our consideration, and $w_j$ is the value of the window function, which 
\begin{equation}
    w_j = \exp\!\left[-\frac{(j - S/2)^2}{(S/4)^2}\right] .
\end{equation}
The pattern speed sampling depends on the total number of snapshot and the time cadence we used, which
\begin{equation}
    \Omega_k = \frac{2\pi}{m}\,\frac{k}{S\,\Delta t} \,,
\qquad k=0,1,\ldots, \frac{S}{2} ,
\end{equation}
where $\Delta t = 0.05~\rm Gyr$ in our setup. The power at the corresponding pattern speed is
\begin{equation}
P(\Omega_k, r) = \frac{1}{W}
\Big( \big|C_{k,m}(r)\big|^2 + \big|C_{S-k,m}(r)\big|^2 \Big)\,,~ k=1,\ldots, \frac{S}{2}-1\,
\end{equation}
where $W$ is the normalisation that $W = S \sum_{j} w_j$.

The resulting spectrograms are shown in Fig.~\ref{appendix:fig:spectrogram} for galaxies with $f_{\rm gas}=\{20\%,~40\%,~60\%,~80\%\}$ from left to right. The ILR, CR, OLR curves in each galaxy are shown in the white dash-dotted, dashed, and dotted lines, respectively. In the $f_{\rm gas}=20\%$ galaxy, the main feature in the spectrogram corresponds to the galactic bar at the inner galaxy, which induces radial migration around the present-day CR and OLR of the galactic bar as discussed in Section~\ref{sec:results}. The power spectrum around the CR curve dominants over the total power around the ILR and OLR of this galaxy, so the CR resonances of the $m=2$ perturbation in this galaxy has a stronger impact to the galaxy. With increasing gas fraction, we find that the power spectrum around the ILR and OLR of the $f_{\rm gas}=40\%~\rm and ~60\%$ galaxies is conspicuously more prominent compared to those in the $f_{\rm gas}=20\%$ galaxy, more similar to the spectrograms of the galaxy model in \citet{Sellwood_Binney2002}. This implies that the role of ILR and OLR becomes more significant with increasing gas fraction. The pattern speed of the perturbation in \citet{Sellwood_Binney2002} (Fig. 10 therein) is mainly around the ILR and corotation resonances, consistent with the feature of kinematic density waves (e.g. \citealt{Kalnajs1973}) and corotating spiral arms (e.g. \citealt{Grand2012}). For the most gas-rich $f_{\rm gas}=80\%$ galaxy, the spectrograms is more irregular compared to its gas-poor analogues. As shown in Eq.~\ref{eqn:heating_migration}, only the corotation resonances ensures a perfectly cold radial migration (i.e. $\Delta J_R=0$ during migration), the increasing strength around the ILR and OLR features in the more gas-rich galaxies are likely the main reasons responsible to the difference in the secular evolution history of these four galaxies. However, as the correlation between the spectrogram of a galaxy and the secular evolution history remains elusive and not yet been quantified. We do not further discuss into this and leave this to future works (Zhang et al. in prep.).

\section{Radial migration of Newborn stars v.s. Pre-assembled disc}\label{appendix::new_vs_old_stars}

\begin{figure*}
    \includegraphics[width = 0.98\textwidth]{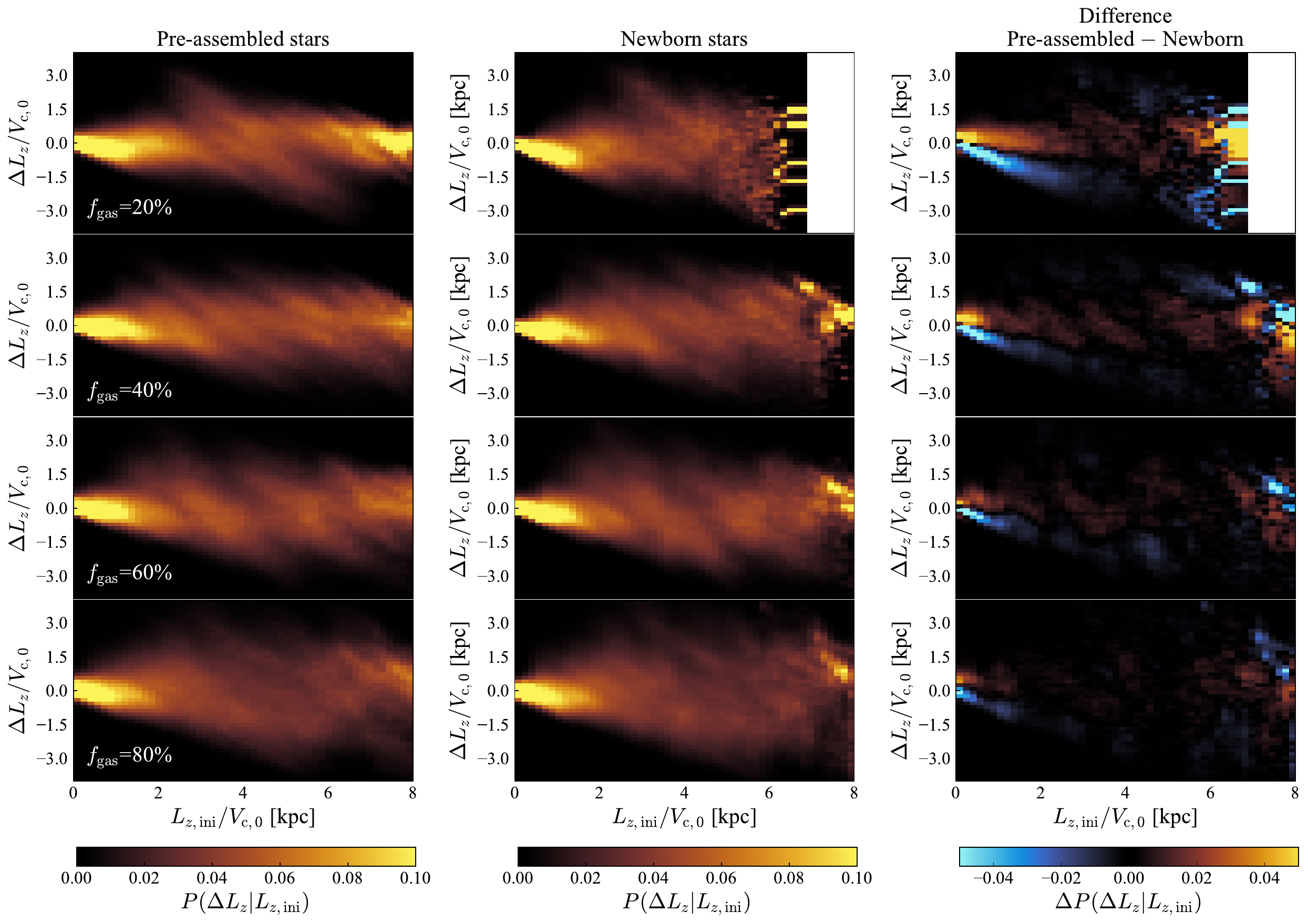}
    \caption{The column-normalised $L_{\rm ini}-\Delta L_z$ plane of pre-assembled disc stars ({\it left}), newborn stars ({\it middle}), and their differences ({\it right}, pre-assembled disc minus newborn stars)}
    \label{appendix:fig:newborn_vs_preborn}
\end{figure*}

The main results of the paper are obtained using pre-assembled disc stars as tracer particles for the secular evolution of galaxies with different gas fractions, whereas newborn stars are not used as tracer particles because their birth radial profile depends on the galaxy's gas fraction. To complete the analysis, we present the radial migration behaviour of the newborn stellar particles and compare it to the pre-assembled disc stars at the same birth radius. We show the column-normalised $L_{\rm ini}-\Delta L_z$ plane of the pre-assembled disc stars and the newborn stars in Fig.~\ref{appendix:fig:newborn_vs_preborn}. The column-normalisation ensures a fair comparison of stars born at the same radius, where the difference in the initial radial profile ($\approx P (L_{\rm ini}/V_c)$) is neglected. To visualise the difference between the pre-assembled stars and the newborn stars, we take their differences and show the residual in the right column of Fig.~\ref{appendix:fig:newborn_vs_preborn}.

Among all four galaxies with different gas fractions, the residuals are relatively minor for most of the initial angular momentum bins except at the innermost and outermost regions of the disc constituted by the newborn stars. For example, at the top right corner, there's a clear residual pattern with $L_{\rm ini}/V_c\lesssim3~\rm kpc$, where the blue region resides in more negative $\Delta L_z$ regions, implying stronger angular momentum loss for newborn stellar particles. This signature suggests that there are more newborn stars trapped in bar orbits after their birth. This could be either a consequence of the radial shear flow of gas, as suggested and discussed more intensively in \citet{BH2024}, or of the kinematic fragmentation, where kinematically colder particles respond more actively to the bar potential \citep[][]{Debattista2017, Fragkoudi2017, Zhang2024}. This difference does not concern the argument of the paper, as we focus on the disc, not the bar. We refer the audience to \citet{BH2024} and \citet{Davis2025} for more discussions on the differences between the inner galaxies of the newborn stars and the pre-assembled discs. The residual at the outermost part of the newborn discs is likely due to low number statistics, as the residual at large $L_{\rm ini}/V_c$ becomes smaller at large $f_{\rm gas}$, where the newborn disc is more spatially extended. In the middle region of the disc, which concerns mostly the argument of this paper, the residuals are minor ($\lesssim10\%$). This suggests that the radial migration behaviour are similar for the pre-assembled disc stars and newborn stars with the same initial angular momentum. Therefore, choosing only the pre-assembled disc stars in our analysis will not affect the argument and conclusion of the paper.

% If you want to present additional material which would interrupt the flow of the main paper,
% it can be placed in an Appendix which appears after the list of references.

%%%%%%%%%%%%%%%%%%%%%%%%%%%%%%%%%%%%%%%%%%%%%%%%%%

% Don't change these lines
\bsp	% typesetting comment
\label{lastpage}
\end{document}